\documentclass[12pt,subeqn,a4paper]{article}

\usepackage{amssymb,amsmath,amsfonts,amsthm, amscd, mathrsfs,helvet}
\usepackage{bm}
\usepackage{graphicx,verbatim}
\usepackage[all]{xy}
\usepackage[dvipsnames]{xcolor}
\usepackage{braket}
\usepackage{enumitem}
\usepackage[hidelinks]{hyperref}
\usepackage{cite}
\usepackage[hmargin={22mm,22mm},vmargin={30mm,30mm}]{geometry}
\usepackage{appendix}
\usepackage{caption}
\usepackage[framemethod=TikZ]{mdframed}
\usepackage{tocloft}

\numberwithin{equation}{section}

\newcommand{\N}{\mathcal{N}}
\newcommand{\M}{\mathcal{M}}

\definecolor{mygreen}{RGB}{43,159,129}
\definecolor{mygreendarker}{RGB}{32,125,101}

\def\r2{\sqrt{2}}

\renewcommand{\baselinestretch}{1.4}

\newcommand{\Z}{\mathcal{Z}}
\newcommand{\hZ}{\hat{\mathcal{Z}}}

\newcommand{\nn}{\nonumber}
\renewcommand{\H}{\mathcal{H}}

\newcommand{\ptt}{\nu}

\newcommand{\I}{\mathcal{I}}
\newcommand{\hI}{\hat{\mathcal{I}}}

\newcommand{\Q}{\mathcal{Q}}
\renewcommand{\S}{\mathcal{S}}

\renewcommand{\r}{L}

\newcommand{\lens}{\mathcal{L}}

\global\mdfdefinestyle{takeaway}{ 
linecolor=mygreen,
linewidth=2pt,
leftmargin=2cm,
rightmargin=2cm, 
roundcorner=10pt,
innertopmargin=8pt,
innerbottommargin=8pt}

\begin{document}
\newcommand{\nd}[1]{/\hspace{-0.5em} #1}
\begin{titlepage}
\begin{flushright}
{\bf March 2023} \\ 
\end{flushright}
\begin{centering}
\vspace{.2in}
 {\Large {\bf Conformal Quantum Mechanics, Holomorphic Factorisation, and Ultra-Spinning Black Holes}}

\vspace{.3in}

Nick Dorey and Rishi Mouland\\
\vspace{.1 in}
DAMTP, Centre for Mathematical Sciences \\ 
University of Cambridge, Wilberforce Road \\ 
Cambridge CB3 0WA, UK \\
{\tt N.Dorey@damtp.cam.ac.uk, r.mouland@damtp.cam.ac.uk} \\
\vspace{.2in}
%
%
\vspace{.4in}
{\bf Abstract} \\[1.5em]
\begin{minipage}{0.91\textwidth}

We study a limit in which a relativistic CFT reduces to conformal quantum mechanics, and relate the partition functions of the two theories. When the initial CFT is holographic, our limit coincides with an ultra-spinning limit in the gravity dual. We therefore propose that ultra-spinning black holes are dual to an appropriate ensemble in finite-dimensional conformal quantum mechanics. The limit is studied in detail for SCFTs in four and six dimensions. These theories have a superconformal index which can be computed by gluing together two or more blocks. Applying our limit to the index effectively isolates a single such block. Our results therefore suggest that ultra-spinning black holes play the role of blocks in the gravitational dual of holomorphic factorisation.

\end{minipage} 
   
\end{centering}

\end{titlepage}

\setcounter{tocdepth}{2}
\setlength\cftbeforesecskip{8pt}
\renewcommand{\baselinestretch}{1.1}\normalsize
\tableofcontents
\renewcommand{\baselinestretch}{1.4}\normalsize

\newpage 
{\noindent \LARGE\bf\color{mygreen}  Introduction}\\	
\addcontentsline{toc}{section}{\color{mygreen}\large Introduction}

\noindent The last few decades has seen the landscape of known interacting conformal field theories (CFTs) grow vastly. Often motivated by string/M-theory constructions, we now recognise the existence of a broad family of CFTs in dimensions $d\le 6$. However, the direct study of such theories---for instance the determination of their spectra of local operators---is in general difficult. Even if a theory admits a Lagrangian description  in some weakly-coupled regime, little information about strong-coupled behaviour can be gleaned directly. One is thus lead to more indirect approaches.

One approach is to leverage any additional structure the CFT might have.
For instance, if the theory is also supersymmetric (an SCFT), then the study of the BPS spectrum is generically tractable. A key tool towards this goal is the \textit{superconformal index} \cite{Kinney,Bhattacharya:2008zy}, which encodes information on the BPS spectrum valid at all couplings. Another key tool in the modern CFT arsenal is holography \cite{Maldacena:1997re}, which for distinguished CFTs offers a dual gravitational description in anti-de-Sitter (AdS) space. Black holes in this dual theory then correspond to an ensemble of states in the CFT. These two approaches are thus neatly united when we study supersymmetric black holes, whose microstates contribute to the superconformal index.

One could however take a rather different perspective; that the full field theory is just too hard, and that we should seek some limit or special background in which the theory simplifies. We can then study this simplified theory, either in its own right, or with an aim to learn something about the theory we started with.
One famous example of such an approach is discrete lightcone quantisation\footnote{DLCQ is potentially much more than a simplifying limit of the theory and it has been conjectured that one can recover the full dynamics of the original theory from DLCQ in an appropriate limit, or else (at least for a CFT) by suitably deforming what we mean by the DLCQ \cite{Lambert:2020zdc,Lambert:2021mnu,Lambert:2021fsl,Lambert:2021nol}. We will not assume this stronger property in the following.} (DLCQ), in which one formulates a theory on a geometry with a compact null circle. For theories in Minkowski space, such a background can be reached as a limit of the same theory on a spacelike circle \cite{Seiberg}. 
The resulting simplification is that sectors of the theory with fixed momentum in the compact direction reduce to finite dimensional quantum mechanics in this limit\footnote{As we review below, generically such an effective description only emerges after integrating out a zero mode sector.}.
As we shall review in more detail shortly, for a conformal theory in radial quantisation the relevant background is the null-compactified pp-wave geometry \cite{Nishida:2007pj,Goldberger:2008vg,Maldacena:2008wh}.
Under favourable circumstances, sectors of fixed null momentum in this geometry are described by finite dimensional conformal quantum mechanics \cite{Fubini}.\\

The main purpose of the present work is to investigate the interplay between the two general approaches outlined above. In the context of the simplification to conformal quantum mechanics mentioned above, we ask the question: what special features do the resulting quantum mechanical theories have when the original model is superconformal and/or has a holographic dual?

As a key tool, we will define and study an analogue of Seiberg's limit \cite{Seiberg} relevant to conformal field theory. In particular, we will show that the null-compactified plane wave and the resulting simplified dynamics can also be obtained as the limit of more conventional spacelike compactifications of the same CFT. In cases where the original field theory has a holographic dual we show that the corresponding limit in the gravitational theory is an {\em ultra-spinning limit}, which has been discussed before in the GR literature. In many cases, black hole solutions in the dual theory go over to new solutions known as {\em ultra-spinning black holes}. Hence, building on the results of \cite{Maldacena:2008wh}, we propose a new holographic duality between 
thermal ensembles in conformal quantum mechanics and ultra-spinning black holes.

If the theory of interest is an SCFT, the relevant limiting theory is superconformal quantum mechanics of the type studied in \cite{Eliezer}\footnote{See \cite{Fedoruk:2011aa} and references therein for a recent review.}. These theories also have a superconformal index counting BPS states \cite{KimLee,SingletonDorey}. It is natural to expect that this index can be obtained as an appropriate limit of the field theory index. We formulate this limit, and study it in a class of examples in four and six dimensions. We find that this correspondence is intimately related to the notion of \textit{holomorphic factorisation}, and in the process uncover a new relation between the 
DLCQ index and the so-called holomorphic blocks appearing in the factorisation of superconformal indices.\\

In the remainder of this introduction, we first provide more detail on the background and motivation for the present work, before providing a detailed summary of our key results.

\newpage 
\section*{Background and motivation}
\addcontentsline{toc}{section}{\hspace{4.6mm} Background and motivation}

Firstly, consider a CFT on Minkowski space $\mathbb{R}^{1,d-1}$. By the usual operator-state map of relativistic conformal field theory, the spectrum of local operators corresponds precisely to the spectrum of states in radial quantisation; that is, we formulate the theory on $\mathbb{R}_t\times S^{d-1}$, and consider states on $S^{d-1}$.
One setting in which we are often able to explicitly compute quantitative facts about the spectrum on $S^{d-1}$ is when the theory is superconformal. We can then define a \textit{superconformal index} \cite{Kinney,Romel,Bhattacharya:2008zy}, which 
receives contributions only from states saturating a particular BPS bound.

An interesting  property of many such indices is that of \textit{holomorphic factorisation}\footnote{The phenomenon of holomorphic factorisation has been observed for a great deal of Euclidean partition functions for theories in various dimensions, on various background geometries; useful reviews of such results can be found in \cite{Nieri:2015yia,Pasquetti:2016dyl}.}. In detail, supersymmetric partition functions on the sphere take the form of an integral over gauge fugacities. The integrand corresponds to a partition function which counts all BPS states, while performing the integral effectively projects onto the gauge singlet sector. Typically the integrand can be written as the product of factors or blocks associated to individual coordinate patches on the sphere. Remarkably the same blocks can be glued together in different ways to obtain supersymmetric partition function on various compact manifolds. This property has its origin in the relation of the resulting blocks to the observables of an associated topological QFT. The factorisation becomes manifest when the partition function is evaluated using localisation and the coordinate patches in question are neighbourhoods containing fixed points of a group action. 
For certain three- \cite{Beem:2012mb} and four-dimensional \cite{Nieri:2015yia} theories a stronger notion of factorisation exists and the full integrated partition function can be decomposed in terms  of integrated blocks. We will make contact with both the strong and weak forms of factorisation in the following.         

A key focus of this paper is the superconformal indices of theories in four and six dimensions, which can be understood as partition functions on $S^1\times S^3$ and $S^1 \times S^5$, respectively. In particular, the indices of four-dimensional $\N=1$ theories admit a ``strong" factorisation into two integrated holomorphic blocks \cite{Nieri:2015yia,Yoshida:2014qwa}, a phenomenon which admits an elegant geometric interpretation as the gluing of partition functions on $T^2 \times D^2$. Relevant also for us is a proposal for the index of the six-dimensional $U(N)$ $(2,0)$ theory in terms of the partition function of a five-dimensional theory on $S^1 \times \mathbb{CP}^2$ \cite{Kim:2012tr,Kim:2013nva}. The latter is then found to localise to the three fixed points of the $U(1)^2$ action on $\mathbb{CP}^2$, and thus the six-dimensional index admits a (weak) factorisation into three five-dimensional holomorphic blocks. This is a particular example of a more general notion of holomorphic factorisation in five dimensions \cite{Pasquetti:2016dyl}. 

When the SCFT is holographic, the dual gravitational theory generically admits BPS black hole solutions, whose entropy must correspond to a large degeneracy of BPS states. Indeed, such a growth of states has been observed in the indices of many such theories, starting with black holes in AdS$_4\times S^7$ \cite{Benini:2015eyy}, followed by those in AdS$_5\times S^5$ \cite{Choi:2018hmj,Benini:2018ywd,Cabo-Bizet:2018ehj}, and more recently for a slew of black objects in various backgrounds\footnote{See \cite{Zaffaroni:2019dhb} for a partial review.}. 

Suppose then we have a holographic SCFT whose index exhibits a holomorphic factorisation.
The realisation of such a factorisation on the gravity side---in particular in the physics of the dual BPS black holes---remains mysterious. Some progress has been made through the observation that BPS black hole entropy\footnote{More precisely, the decomposition is realised by the entropy functionals introduced in \cite{Hosseini:2017mds,Hosseini:2018dob}, whose constrained Legendre transform reproduces the Bekenstein-Hawking entropies of BPS black holes in various dimensions.} in many cases admits a decomposition into ``gravitational blocks" \cite{Hosseini:2019iad,Hosseini:2021mnn}, suggesting that some independent factorisation of gravitational degrees of freedom could be possible. However, the question still remains: is there a black hole whose microstates are precisely those states counted by a \textit{single} holomorphic block? One result of this paper is that the answer is \textit{yes}, and that these black holes are of so-called ``ultra-spinning" type.\\

Secondly, consider the same CFT but now formulated on a null compactification of Minkowski space; i.e. in discrete lightcone quantisation (DLCQ). The compactification breaks the full conformal group down to the Schr\"odinger group; the resulting theory is thus a non-relativistic conformal field theory (NRCFT). One may then want to study the spectrum of local\footnote{Here, a local operator is defined as an operator with fixed (discrete) momentum on the null circle, while being local in the remaining directions.} operators. The NRCFT operator-state map \cite{Nishida:2007pj} then provides a precise correspondence between the spectrum of local operators and the spectrum of states of the theory placed on the null-compactified pp-wave spacetime \cite{Duval:1994qye,Goldberger:2008vg}. The examples we will focus on are superconformal, with the null compactification preserving some supersymmetry; one is then able to define a Witten index that counts states on the pp-wave saturating a particular BPS bound. 

In principle, the DLCQ of a CFT in a sector of fixed momentum along the null circle is a finite-dimensional conformal quantum mechanics. The operator state map described above can then be understood as a reorganisation of the spectrum in each of these quantum mechanics into discrete eigenstates of a particular Hamiltonian \cite{SingletonThesis}. However, for a generic CFT the explicit determination of these quantum mechanical models is a notoriously subtle task. As pointed out in \cite{Hellerman:1997yu}, the subtlety lies in correctly accounting for the effect of modes carrying zero momentum on the compact null circle; indeed, dealing with the analogous modes in lightcone quantisation was the key motivation in defining DLCQ in the first place \cite{Maskawa:1975ky}. This issue can be circumvented for some SCFTs in four and six dimensions whose string-theoretic interpretation gives rise to explicit proposals for their DLCQ quantum mechanics \cite{Aharony:1997th,Aharony:1997an,Ganor:1997jx,Kapustin:1998pb}. Note however that a key result of this paper is a procedure by which one can recover the partition function of the DLCQ theory as a limit of partition functions computed in radial quantisation\footnote{More precisely, computed on the lens space, as we will see.}. In other words, this paper offers an alternative means to study the spectrum of the putative DLCQ theory that does not require an explicit construction of the effective conformal quantum mechanics.

If the CFT is holographic, then the DLCQ theory has dual description as a gravitational theory formulated on asymptotically null-compactified AdS space. In suitable coordinates, such spacetimes realise as their conformal boundary the null-compactified pp-wave. Dualities of this form were first studied in \cite{Son:2008ye,Balasubramanian:2008dm,Goldberger:2008vg,Barbon:2008bg,Adams:2008wt,Herzog:2008wg,Maldacena:2008wh}. Black hole solutions in various gravity theories with precisely these asymptotics were first studied in \cite{Maldacena:2008wh}; for reasons that will soon become clear, we will refer to them as \textit{ultra-spinning black holes}.
A key observation of \cite{Maldacena:2008wh} is that one requires the black hole to have sufficiently large lightcone momentum in order to have a semiclassical approximation to gravity in the bulk.

With such an ultra-spinning black hole in hand, one can compute its Bekenstein-Hawking entropy, and seek to match it against the leading-order growth of states in the DLCQ theory with the same charges as the black hole. For the case of seven-dimensional ultra-spinning supersymmetric black holes, precisely such a match was achieved in \cite{Dorey:2022cfn}. The relevant conformal quantum mechanics arises from the DLCQ of the six-dimensional $U(N)$ $(2,0)$ theory, and in each subsector of fixed particle number $K$ is realised as precisely the superconformal quantum mechanics of $K$ Yang-Mills instantons of $SU(N)$. In \cite{Dorey:2022cfn} the degeneracies of BPS states in this quantum mechanics were probed through an asymptotic study of the superconformal index, and shown to precisely reproduce the Bekenstein-Hawking entropy of the seven-dimensional ultra-spinning black hole solutions presented there.


\section*{Summary of results}
\addcontentsline{toc}{section}{\hspace{4.6mm}  Summary of results}

As mentioned above, a key tool in concretely formulating the DLCQ of a generic theory in Minkowski space is the limiting procedure put forward in \cite{Seiberg}, by which one \textit{defines} the DLCQ as a particular limit of spatial compactifications. This allows one to keep track of what sectors of the spectrum of the original theory survive in the DLCQ. In the context of conformal field theory, the relevant spectral problem in the original theory sees us formulate it in radial quantisation, while the DLCQ theory is formulated on the null-compactified pp-wave. The aim of {\color{mygreendarker}\textbf{Part I}} is to define and study an analogue of Seiberg's limit applicable to this conformal setting, which takes us between these two quantisations.

The limit we consider is based on the following geometric considerations. It is well known that the Lorentzian spacetime $\mathbb{R}_t \times S^{d-1}$ admits a \textit{Penrose limit}, in which one zooms in near a null geodesic that wraps some circle in the sphere. The resulting geometry is the pp-wave spacetime. In order to arrive at the DLCQ geometry, we instead start with an orbifold $\mathbb{R}_t \times (S^{d-1}/\mathbb{Z}_L)$. Our main focus is on even-dimensional CFTs, for which this orbifold can be chosen to be free; the resulting spaces $S^{d-1}/\mathbb{Z}_L$ are precisely lens spaces. Then, by taking a coordinated Penrose and $L\to\infty$ limit, one arrives at the desired null-compactified pp-wave.

This motivates a relationship between a CFT formulated on lens space, and in DLCQ. For this, it is key to recognise that in this limit, slices of constant time in the starting geometry $\mathbb{R}_t \times (S^{d-1}/\mathbb{Z}_L)$ go over to those of constant lightcone time in the DLCQ. The theory on $\mathbb{R}_t \times (S^{d-1}/\mathbb{Z}_L)$ can be obtained by orbifolding the theory on the sphere, which in a gauge theory generically introduces new twisted sectors corresponding to discrete holonomies on the lens space. Then, let $\hat{\Z}_L(\hat{\nu})$ denote the refined partition function of the lens space theory, defined as a trace of states on $S^{d-1}/\mathbb{Z}_L$, and parametrised by the inverse temperature and some chemical potentials, collectively $\{\hat{\nu}\}$. Similarly, we can define a refined partition function $\Z(\nu)$ of the DLCQ theory, again with some inverse temperature and chemical potentials $\{\nu\}$. When a manifest conformal quantum mechanical description is available in each sector of fixed null momentum, $\Z(\nu)$ can be computed as a sum over the partition functions of these theories. We then propose a quantitative relationship of the form
\begin{align}
  \Z(\nu) = \lim_{L\to\infty} \hat{\Z}_L\!\left(\hat{\nu}(\nu,L)\right)
  \label{eq: pf limit}
\end{align}
where the parameters $\hat{\nu}$ are fixed as a function of the $\nu$ and the lensing degree $L$ as we take the limit. Note, in a gauge theory we expect such a relationship to hold only when one can avoid the condensation of twisted sector states. We are able to probe this possibility quantitatively for four-dimensional $\N=1$ theories.

We then consider the gravitational perspective on this procedure in the case that the CFT admits a dual holographic description. Finite temperature ensembles in the DLCQ theory are dual to black holes with null-compactified AdS asymptotics. Such black holes can be found by running the same argument above in the bulk: we take a Penrose limit of a known asymptotically AdS black hole, along with a particular coordinate identification. We point out that this limit, first considered in \cite{Maldacena:2008wh}, coincides precisely with the so-called \textit{ultra-spinning limit} of rotating AdS black holes\footnote{See also \cite{Emparan:2003sy,Caldarelli:2008pz,Caldarelli:2011idw} who study a similar limit, whose relation to the limit we're interested in is discussed in Section \ref{sec: BHs} following \cite{Klemm:2014rda}.} \cite{Gnecchi:2013mja,Klemm:2014rda,Hennigar:2014cfa,Hennigar:2015cja}. We discuss in some depth the construction, features and existing literature on such \textit{ultra-spinning black holes}. The key takeaway is:\vspace{0.5em}
\begin{mdframed}[style=takeaway]\centering
	An ultra-spinning black hole is dual to a thermal ensemble in conformal quantum mechanics.
\end{mdframed}
In {\color{mygreendarker}\textbf{Part II}}, we turn our attention to some examples, which share some common features. To probe the quantitative relationship (\ref{eq: pf limit}), we focus on superconformal field theories (SCFTs). Then, provided that the orbifold used to define the lens space theory preserves at least one supercharge, we can specialise partition functions $\hat{\Z}_L$ and $\Z$ to become supersymmetric indices $\hat{\I}_L$ and $\I$ for the lens space and DLCQ theories, respectively. Note, $\I$ can be equivalently regarded as a sum over the superconformal indices of the superconformal quantum mechanics at each fixed lightcone momentum. Then, (\ref{eq: pf limit}) implies a relation of the form
\begin{align}
  \I(\nu) =\lim_{L\to\infty}\hat{\I}_L (\hat{\nu}(\nu,L))
  \label{eq: index limit}
\end{align}
under a precise scaling limit of the parameters\footnote{Necessarily, the $\{\hat{\nu}\}$ and the $\{\nu\}$ are constrained to a codimension-1 subspace, which promotes the respective partition functions to indices.} $\hat{\nu}$.

We will study in detail the six-dimensional $(2,0)$ theory, and generic $\N=1$ theories in four dimensions. As reviewed above, the supersymmetric indices in both of these settings exhibit various notions of holomorphic factorisation. This factorisation is shown to persist in the lens space indices $\hI_L$ for all $L$, which generically also involve a sum over twisted sectors. Then, in all our examples, the limit (\ref{eq: index limit}) is one in which a single block in the integrand dominates as we approach $L\to\infty$. There are a number of theory-dependent features and details one must address in taking the limit precisely, including the fate of the twisted sector contributions. We leave these details to the main text, and for now describe the result. For the six-dimensional $A_{N-1}$ $(2,0)$ theory, and a broad class of four-dimensional $\N=1$ theories, the DLCQ index $\I$ is given as an integral over gauge fugacities, with an integrand that can be related to a holomorphic block of the original theory. In all, we arrive at the following proposal:\vspace{0.5em}
\begin{mdframed}[style=takeaway]\centering
	The DLCQ index $\I$ is given by an integrated holomorphic block.
\end{mdframed}
Note, in the case of the six-dimensional $A_{N-1}$ $(2,0)$ theory, we have available an independent definition of the DLCQ theory at fixed lightcone momentum $K$ as the superconformal quantum mechanics of $K$ Yang-Mills instantons of $SU(N)$; the superconformal indices of the quantum mechanical theories exactly reproduce the lightcone index $\I$ found from $\hat{\I}_L$, thus providing an independent verification of (\ref{eq: pf limit}).

We finally study the gravitational implications of our findings when the SCFT is holographic. Known ultra-spinning black holes in five and seven dimensions admit BPS limits, in which they become both supersymmetric and extremal. Hence, at least in these settings we are led to propose:\vspace{0.5em}

\begin{mdframed}[style=takeaway]\centering
	The microstates of a BPS ultra-spinning black hole are captured by an integrated holomorphic block.
\end{mdframed}
In particular, this suggests that ultra-spinning black holes provide a concrete holographic realisation of the gravitational blocks proposed in \cite{Hosseini:2019iad,Hosseini:2021mnn}.

\section*{Acknowledgements}

We would like to thank Kimyeong Lee, Dario Martelli and Sam Crew for useful discussions.
We would also like to thank Chiung Hwang for collaboration in an early stage of this project.
R.M. was supported by David Tong's Simons Investigator Award.
This work has been partially supported by STFC consolidated grant ST/T000694/1.

\section*{A note on notation}

In this paper we study the same $d$-dimensional CFT in two different quantisations, and determine how they are related. Before long, we will specialise to even $d=2(m+1)$. Then, the following kinds of indices will always run over the specified ranges:
\begin{align}
  i,j,\dots &\in \{1,\dots,2m\}		\nn\\ 
  \mu,\nu,\dots &\in \{1,\dots,m+1\}		\nn\\ 
  \alpha,\beta,\dots &\in \{1,\dots,m\}	
\end{align}
We will first consider lens space quantisation, whereby the theory is formulated on $\mathbb{R}_t \times (S^{2m+1}/\mathbb{Z}_L)$. The partition function $\hZ_L$ along with chemical potentials on which it depends will always be hatted. We will then consider the discrete lightcone quantisation (DLCQ). Then, the partition function $\Z$ along with its chemical potentials will not be hatted. 


\newpage
{\noindent \LARGE\bf\color{mygreen}  Part I: Non-relativistic limits of CFTs}	
\addcontentsline{toc}{section}{\color{mygreen}\large Part I: Non-relativistic limits of CFTs}

\section{The lens space theory}\label{sec: lens space theory}
Let us first consider a $d$-dimensional CFT formulated on flat Minkowski space. We would like to study the spectrum of local operators in the theory, which by the standard relativistic operator-state map amounts to studying the spectrum of states in radial quantisation; that is, putting the theory on $\mathbb{R}_t\times S^{d-1}$ and considering the Hilbert space on $S^{d-1}$. 

Let us next generalise this picture, and consider the theory on the orbifold $\mathbb{R}_t\times (S^{d-1}/\mathbb{Z}_L)$ for some $L\in \mathbb{N}$. Let us further require that the orbifold is free, which requires even $d=2(m+1)$. The corresponding orbifolds are precisely lens spaces.  In detail, let us choose coordinates on $\mathbb{R}_t\times S^{2m+1}$ pulled back from the Minkowski metric on $\mathbb{R}^{1,2m+2}$. These are $(\hat{t},\hat{n}_\mu,\hat{\phi}_\mu)$, where for each $\mu=1,\dots,m+1$, we have polar coordinates $(\hat{n}_\mu,\hat{\phi}_\mu)$ on an $\mathbb{R}^2$, and so $\hat{n}_\mu\ge 0$ and $\hat{\phi}_\mu \sim \hat{\phi}_\mu + 2\pi$ . The metric is
\begin{align}
  \hat{ds}^2 = -d\hat{t}^2 + \sum_{\mu=1}^{m+1} \left(d\hat{n}_\mu^2 + \hat{n}_\mu^2 d\hat{\phi}_\mu^2\right) 
\label{eq: R x S metric}
\end{align}
Then, $\mathbb{R}_t\times S^{2m+1}$ is the submanifold defined by $\sum_\mu \hat{n}_\mu^2 = 1$. Topologically, we are viewing $S^{2m+1}$ as a $T^{m+1}$ fibration over the interior of a standard $m$-simplex. This torus degenerates to $T^{n+1}$ on any $n$-face of this $m$-simplex. 

For reference later, we denote by $\hat{H}$ and $\hat{J}_\mu = (\hat{J}_\alpha, \hat{J}_{m+1})$ the Hermitian generators of translations along $\hat{t}$ and $\hat{\phi}_\mu = (\hat{\phi}_\alpha, \hat{\phi}_{m+1})$, respectively. Bosons have all integer eigenvalues under the $\hat{J}_\mu$, while fermions have all half-integer eigenvalues.

Then for each $L\in \mathbb{N}$, we have a free $\mathbb{Z}_L$ action on $S^{2m+1}$ for each set of integers\footnote{The lens spaces defined in this way are not in general homeomorphically distinct from one another.} $(p^L_{1},p^L_{2},\dots,p^L_{m})$ such that $p^L_{\alpha}$ is coprime to $L$ for all $\alpha=1,\dots,m$. This $\mathbb{Z}_L$ is generated by the element
\begin{align}
  (\hat{n}_\alpha,\hat{n}_{m+1}, \hat{\phi}_\alpha,\hat{\phi}_{m+1}) \quad \longrightarrow \quad \left(\hat{n}_\alpha,\,\,\hat{n}_{m+1}, \,\,\hat{\phi}_\alpha + \frac{2\pi p_\alpha^L}{L},\,\,\hat{\phi}_{m+1} + \frac{2\pi}{L}\right)
  \label{eq: lens space orbifold}
\end{align}
Quotienting by this $\mathbb{Z}_L$ action defines the lens space $S^{2m+1}/\mathbb{Z}_L[p_\alpha^L]$. The simplest example takes $p_1^L=\dots= p_m^L = 1$; in this case, we are simply quotienting along the Hopf fibre $S^1 \hookrightarrow S^{2m+1} \rightarrow \mathbb{CP}^m$.

\subsection{Theories on lens space via orbifolding}

Let us now discuss how we arrive at a theory on $\mathbb{R}_t\times S^{2m+1}/\mathbb{Z}_L[p_\alpha^L]$ by taking an appropriate orbifold of the theory on $\mathbb{R}_t\times S^{2m+1}$. Many lens space theories studied in the literature---both in four \cite{Benini:2011nc,Alday:2013rs,Razamat:2013jxa,Razamat:2013opa} and six \cite{Kim:2012tr,Kim:2013nva} dimensions---provide examples of this more general construction.

\subsubsection{Bosonic theories}

For a theory with only bosonic degrees of freedom, we orbifold the theory by the action of a spacetime rotation $g_L\in SO(2m+2)$ on the sphere, where 
\begin{align}
  g_L=\exp\left(\frac{2\pi i}{L}\left(p_\alpha^L \hat{J}_\alpha + \hat{J}_{m+1} \right)\right)
\end{align}
In particular, each $\hat{J}_\mu$ has integer eigenvalues, and thus $(g_L)^L=1$ on all states. Quotienting by the action of $g_L$ thus defines a $\mathbb{Z}_L$ spacetime orbifold. The orbifold projects onto states invariant under $g_L$, i.e. those satisfying
\begin{align}
  p_\alpha^L \hat{J}_\alpha + \hat{J}_{m+1} \in L\mathbb{Z}
  \label{eq: orbifold quantisation} 
\end{align}
This in effect imposes periodicity conditions on all fields of the theory.

With the orbifold projection performed one must subsequently in general introduce additional `twisted' sectors that were not present in the original theory.
In particular, in a gauge theory, the non-trivial fundamental group $\pi_1(S^{2m+1}/\mathbb{Z}_L[p^L_\alpha])=\mathbb{Z}_L$ allows for non-trivial gauge bundles, characterised by non-zero (but discrete) holonomies along the non-contractible cycles; the full Hilbert space on $S^{2m+1}/\mathbb{Z}_L[p^L_\alpha]$ must include sections over such bundles\footnote{A concrete analysis of such bundles and their sections is performed for $S^3$---relevant for the four-dimensional case $m=2$---in \cite{Alday:2012au}.}. For this reason, when it comes to our main focus of counting states contributing to a partition function, in a gauge theory one cannot in general simply throw away states not satisfying (\ref{eq: orbifold quantisation}) and get the partition function of the lens space theory.

\subsubsection{Including fermions}

The situation is slightly more subtle when the theory has fermions. Once again, we want to orbifold our theory by the action of 
\begin{align}
  g_L=\exp\left(\frac{2\pi i}{L}\left(p_\alpha^L \hat{J}_\alpha + \hat{J}_{m+1} \right)\right)
\end{align}
where now $g_L$ generates a cyclic subgroup of $\text{Spin}(2m+2)$. This subgroup is $\mathbb{Z}_{2L}$ rather than $\mathbb{Z}_L$ when $(\sum_\alpha p_\alpha^L)$ is even, since
\begin{align}
  (g_L)^L = (-1)^{(\sum_\alpha p^L_\alpha+1)F}
  \label{eq: internal orbifold condition}
\end{align}
for fermion number operator $F$ (concretely, we could take for instance $F=2\hat{J}_1$).

Thus, for odd $(\sum_\alpha p_\alpha^L)$, $g_L$ generates a $\mathbb{Z}_L$ spacetime orbifold, and we can proceed as before by imposing particular periodic boundary conditions on fields, and introducing any twisted sectors. In particular, the $L=1$ case is trivial and we retain the original theory.
 
 In contrast, when $(\sum_\alpha p_\alpha^L)$ is even we have $(g_L)^L=(-1)^F$ and thus the orbifold projects out entirely all fermionic degrees of freedom, leaving a purely bosonic theory\footnote{For a two-dimensional theory on $\mathbb{R}_t\times S^1$, i.e. $m=0$, one can retain fermions under this orbifold by allowing for anti-periodic (i.e. Neveu-Schwarz) boundary conditions. Our main focus will be theories in four or more dimensions ($m\ge 1$), where the simply-connectedness of $S^{2m+1}$ means that no analogous boundary conditions are possible.}; this is true even at $L=1$. For $L>1$, one can then proceed as before, performing the spacetime orbifold by $g_L$, which now generates a $\mathbb{Z}_L$ action on the bosonic theory.\\
 
In summary, the orbifold takes the form
\begin{align}
  (\mathbb{Z}_L)^\text{spacetime} \times (\mathbb{Z}_k)^\text{internal}
\end{align}
where $k=1$ or $2$ for $(\sum_\alpha p_\alpha^L)$ odd or even, respectively. In the former case, the internal orbifold is trivial and we have only a $\mathbb{Z}_L$ spacetime orbifold; we refer to such an orbifold as \textit{good}\footnote{Note, our use of the nomenclature ``good'' and ``bad'' with regard to orbifolds is entirely distinct from the common use in the mathematics literature, in which a good orbifold is just any orbifold which arises as a global quotient by a discrete group action.}. In the latter case, the internal orbifold is non-trivial, and thus a class of states (in this case, all fermions) are projected out; we refer to such an orbifold as \textit{bad}.

It will turn out that in order to recover the lightcone partition function in a limit of the lens space partition function, we will require that the lens space theory is found as a \textit{good} orbifold of the original CFT. Thus, it appears as though we require $(\sum_\alpha p_\alpha^L)$ odd. In order to get around this and define \textit{good} orbifolds for even $(\sum_\alpha p_\alpha^L)$, we need to consider twisting $g_L$ by internal symmetries; indeed, this will prove crucial when we turn to a six-dimensional example in Section \ref{sec: 6d}.

\subsubsection{Generic twisting}

We generalise by dressing the orbifold generator by a general twist by global symmetries. Let us denote by $\{Q_a\}$ the Cartan generators of any global symmetries of the theory. To get the most general twist by internal symmetries, we add the option of a further sign twist dictated by fermion number. The resulting operator to consider takes the form
\begin{align}
  g_L(n,\alpha) = (-1)^{nF}\exp\left(\frac{2\pi i}{L}\alpha_a Q_a\right) g_L = (-1)^{nF} \exp \left(\frac{2\pi i}{L}\left(p_\alpha^L \hat{J}_\alpha + \hat{J}_{m+1} + \alpha_a Q_a\right)\right)
  \label{eq: general orbifold generator}
\end{align}
where $n=0,1$, and the $\alpha_a$ are some real numbers.

The resulting orbifold takes the form 
\begin{align}
  (\mathbb{Z}_L)^\text{spacetime} \times (\mathbb{Z}_k)^\text{internal}
\end{align}
where now $k\in \mathbb{N} \cup \{\infty\}$, with $(\mathbb{Z}_k)^\text{internal}$ generated by
\begin{align}
  h(n,\alpha) = (-1)^{(\sum_\alpha p^L_\alpha+nL+1)F}\exp\!\big(2\pi i\alpha_a Q_a\big)
\end{align}
If $h(n,\alpha)=1$ on all states in the initial theory on $\mathbb{R}_t\times S^{2m+1}$, then $k=1$ and we have a \textit{good} orbifold. Otherwise, the orbifold is \textit{bad}.

\subsection{A refined partition function}

Given our theory now defined on $\mathbb{R}_t\times S^{2m+1}/\mathbb{Z}_L[p^L_\alpha]$, whose form depends implicitly on our choice of the orbifold parameters $(n,\alpha)$, we can now write down the most general refined partition function. We define the \textit{lens space partition function}
\begin{align}
  \hZ_L \!\left(\hat{\mu},\hat{w}_\mu,\hat{u}_a\right) = \text{Tr}_{L}\exp \Big[-\hat{\mu} \hat{H} - \hat{w}_\mu \hat{J}_\mu - \hat{u}_a Q_a \Big]
\label{eq: lens pf}
\end{align}
Here, the trace is taken over the Hilbert space of states on $S^{2m+1}/\mathbb{Z}_L[p^L_\alpha]$, and as above, the $Q_a$ are Cartan generators for any global symmetries the theory may have.


\section{The lightcone theory}\label{sec: lightcone theory}

Let us now consider the same CFT in $d=2m+2$ dimensions in discrete lightcone quantisation (DLCQ); that is, on Minkowski space
\begin{align}
  \tilde{ds}^2 = -2 d\tilde{x}^+ d\tilde{x}^- + d\tilde{x}^i d\tilde{x}^i
  \label{eq: Minkowski}
\end{align}
with $i=1,\dots,2m$, with the null coordinate $\tilde{x}^+$ identified as $\tilde{x}^+ \sim \tilde{x}^+ + 2\pi$. The full group of conformal symmetries of Minkowski space is broken by this compactification to the centraliser of translations along $\tilde{x}^+$; we have $SO(2,d) \to \text{Schr}(d-2)$, the Schr\"odinger group in $d-2=2m$ spatial dimensions. The reduction of the theory into modes along the null circle is therefore a non-relativistic conformal field theory (NRCFT), with time $\tilde{x}^-$ and spatial directions $\tilde{x}^i$. The `particle number' central charge $K$ is identified with the discrete null momentum along $\tilde{x}^+$. Full details of the rest of the symmetry algebra can be found in Appendix \ref{app: algebra}; let us for now just point out an $SO(2,1)=SL(2,\mathbb{R})$ subgroup generated by Hamiltonian $H$, Lifshitz dilatation $D$, and special conformal generator $C$.
 
\subsection{States on a plane}

We once again want to study the spectrum of local operators in the NRCFT. By the standard operator-state map in this context \cite{Nishida:2007pj}, local\footnote{From the perspective of the $d$-dimensional theory, a local operator is defined as an operator with fixed (discrete) momentum on the null circle, while being local in the remaining directions.} operators of definite scaling dimension under the Lifshitz scaling symmetry $D:(\tilde{x}^-,\tilde{x}^i)\to (\lambda^2\tilde{x}^-,\lambda\tilde{x}^i)$ are in one-to-one correspondence with eigenstates of the so-called \textit{oscillator Hamiltonian} $\Delta:=H+C$. This arises due to an inner automorphism on the (complexified) Schr\"odinger algebra, that maps $-iD\to \Delta$.

As first pointed out in \cite{Duval:1994qye} (see also \cite{Goldberger:2008vg}), this map admits a nice geometric interpretation. Let us define new coordinates $(x^+, x^-, x^i)$ by
\begin{align}
  x^+ 	&= \tilde{x}^+ - \frac{1}{2} \frac{\tilde{x}^i \tilde{x}^i \tilde{x}^-}{1+ (\tilde{x}^-)^2} ,\nn\\
  x^- 	&= \arctan\left(\tilde{x}^-\right),\nn\\
  x^i 	&= \frac{\tilde{x}^i}{\sqrt{1+ (\tilde{x}^-)^2}}
\end{align}
Note, the periodicity of $\tilde{x}^+$ translates simply to $x^+\sim x^+ + 2\pi$, while constant time slices in our original frame ($\tilde{x}^-=\text{const}$) are mapped to constant time slices in the new frame ($x^-=\text{const}$). Then, in the new frame, we have simply that $\Delta$ is the Hamiltonian: it generates translations in the new time $x^-$. Furthermore, $K$ is simply the momentum along the $x^+$ circle.

The metric now takes the form $\tilde{ds}^2 = \sec^2( x^-) ds^2$, where
\begin{align}
  ds^2 = - 2 dx^+ dx^- -  x^i x^i ( dx^-)^2 + dx^i dx^i
  \label{eq: pp-wave}
\end{align}
Hence, after a Weyl rescaling (to which our CFT is blind\footnote{In principle our CFT may have a non-vanishing Weyl anomaly, but since Minkowski space is Ricci-flat, the Weyl rescaling goes through anyway.}), we have precisely the pp-wave spacetime $ds^2$ subject to the null compactification $x^+\sim x^+ + 2\pi$.

We learn therefore that operators of definite Lifshitz scaling dimension on the background (\ref{eq: Minkowski}) are in one-to-one correspondence with energy eigenstates of the theory on the null-compactified pp-wave spacetime (\ref{eq: pp-wave}), where $x^-$ plays the role of time.

\subsection{Another refined partition function}

We can then once again define the most general refined partition function for the theory in DLCQ. We define the \textit{lightcone partition function}
\begin{align}
  \Z(\beta,\mu,w_\alpha,u_a) = \text{Tr}_{\text{LC}} \exp \Big[ - \beta K - \mu \Delta - w_\alpha J_\alpha - u_a Q_a \Big]
  \label{eq: lc pf}
\end{align}
where here the trace is taken over the Hilbert space formulated on constant $x^-$ slices of the null-compactified pp-wave spacetime (\ref{eq: pp-wave}). The full set of Cartan generators are made up of the discrete null momentum $K$, Hamiltonian $\Delta$ generating $x^-$ translations, commuting rotations $J_\alpha$, $\alpha = 1,\dots,m$ in the $x^i$ directions, and once again the $Q_a$ generate any global symmetries. Unitarity ensures that all states have $K\ge 0$, while all non-vacuum states have $\Delta\ge (d-2)/2=m$.

\section{From lens to lightcone}\label{sec: lens to lc}

Let us now understand how the lens space partition function $\hZ_L$ and lightcone partition function $\Z$ are related.

\subsection{A limit of geometries}

First note that $\mathbb{R}_t\times S^{2m+1}$ admits a limit---known as the Penrose limit---in which it becomes precisely the pp-wave spacetime (\ref{eq: pp-wave}) without any null compactification, with slices of constant time $\hat{t}$ going over to slices of constant $x^-$. The key observation is that by instead starting with $\mathbb{R}_t\times S^{2m+1}/\mathbb{Z}_L[p^L_\alpha]$ and also taking $L\to\infty$ as we take the Penrose limit, we arrive at the null-compactified pp-wave.

In detail, let us choose new coordinates $(x^+, x^-, n_\alpha,\phi_\alpha)$ on $S^{2m+1}/\mathbb{Z}_L[p_\alpha^L]$, where recall $\alpha=1,\dots,m$, defined by
\begin{align}
  x^+ = \eta^2 (\hat{t} + \hat{\phi}_{m+1}),\quad x^- = \hat{t},\quad n_\alpha = \eta \hat{n}_\alpha, \quad \phi_\alpha = \hat{\phi}_\alpha
  \label{eq: Penrose limit coordinate transformation}
\end{align}
The coordinates $n_\alpha$ satisfy $n_\alpha \ge 0$ and $\sum_\alpha n_\alpha^2 \le \eta^2$. Further, $x^+$ is periodic with $x^+ \sim x^+ + 2\pi \eta^2$, while of course $\phi_\alpha \sim \phi_\alpha + 2\pi$. The $\mathbb{Z}_L$ orbifold identifies points as
\begin{align}
  \left(x^+, x^-, n_\alpha, \phi_\alpha\right) \quad \sim \quad \left(x^+ + \tfrac{2\pi \eta^2}{L}, x^-, n_\alpha, \phi_\alpha  + \tfrac{2\pi p_\alpha^L}{L}\right)
  \label{eq: finite L identification}
\end{align}
The Penrose limit is simply $\eta\to\infty$, where $x^+$ decompactifies, the coordinates $n_\alpha\ge 0$ become non-compact, and we end up with the pp-wave spacetime (\ref{eq: pp-wave}) up to a constant Weyl factor. But let us instead take both $\eta,L\to\infty$. More precisely, we fix
\begin{align}
  \eta = \sqrt{L}
\end{align}
and take the limit $L\to\infty$, assuming that the integers $p_\alpha^L$ are chosen such that $\lim_{L\to\infty}(p_\alpha^L/L) = 0$ for each $\alpha$. In this limit, we find the following. Firstly, the periodicity $x^+\sim x^+ + 2\pi L$ arising from the definition of $x^+$ in terms of $\hat{\phi}_{m+1}$ is removed, while the coordinates $n_\alpha\ge 0$ become non-compact. Secondly, we have
\begin{align}
  ds^2 := \eta^2 \hat{ds}^2 = -2 dx^+ dx^- -  \rho^2 (dx^-)^2 + \sum_{\alpha=1}^m \left(dn_\alpha^2 + n_\alpha^2 d\phi_\alpha^2\right) + \mathcal{O}(\eta^{-2})
\end{align}
where $\rho^2 = \sum_\alpha n_\alpha^2$. Hence, in the limit the Weyl-rescaled metric $ds^2$ becomes simply
\begin{align}
  ds^2 \,\,\longrightarrow\,\, -2 dx^+ dx^- -  \rho^2 (dx^-)^2 + \sum_{\alpha=1}^m \left(dn_\alpha^2 + n_\alpha^2 d\phi_\alpha^2\right)
\end{align}
which is precisely the pp-wave metric (\ref{eq: pp-wave}), with $m$ sets of polar coordinates parameterising the $x^i$ directions. Finally, in the limit the $\mathbb{Z}_L$ orbifold acts only on the $x^+$ coordinate, on which it simply imposes the periodicity $x^+ \sim x^+ + 2\pi$. Hence, we find precisely the null-compactified pp-wave, as desired. Furthermore, we see that slices of constant $\hat{t}$ do indeed go over to slices of constant $x^-$ in the pp-wave, implying not only a limit of geometries, but also a limit of quantisations (i.e. choice of foliation of spacetime).

\subsection{A limit of partition functions}

The above geometric limit implies that the lightcone partition function $\Z$ as defined in (\ref{eq: lc pf}) can be found in a limit of the lens space partition function $\hZ_L$ as defined in (\ref{eq: lens pf}), provided that the lens space theory is defined with respect to a suitable orbifold.

In particular, it is clear that we require a \textit{good} orbifold, since we do not want to quotient by any internal symmetries, so let us assume this and also take\footnote{It would be interesting to consider a sequence of \textit{good} orbifolds with $n=1$. The result in the $L\to\infty$ limit would be the DLCQ but with anti-periodic boundary conditions for fermions on the null circle.} $n=0$ for all $L$. The Penrose limit is rephrased as a necessary rescaling of chemical potentials as we take $L\to\infty$. These asymptotics can be deduced by recasting the Cartan generators $\hat{H},\hat{J}_\mu$ in terms of momenta $(K,\Delta,J_\alpha)$ in the $(x^+, x^-, \phi_\alpha)$ directions at finite $\eta = \sqrt{L}$; we have
\begin{align}
  K = \frac{1}{L} \hat{J}_{m+1},\quad \Delta = \hat{H} - \hat{J}_{m+1},\quad J_\alpha = \hat{J}_\alpha
  \label{eq: field theory charge rescaling}
\end{align}
Hence, using (\ref{eq: lens pf}) and (\ref{eq: lc pf}), we are lead to propose
\begin{align}
  \lim_{L\to\infty} \hat{Z}_L(\hat{\mu},\hat{w}_\mu,\hat{u}_a) = \Z(\beta,\mu,w_\alpha,u_a)
  \label{eq: general pf limit}
\end{align}
where
\begin{align}
  \hat{\mu} 		&= \mu + \frac{\rho}{L} +  \mathcal{O}\!\left(L^{-2}\right),\nn\\
   \hat{w}_\alpha 	&= w_\alpha +\mathcal{O}\!\left(L^{-1}\right),\nn\\
    \hat{w}_{m+1} 	&= -\mu + \frac{\beta-\rho}{L} + \mathcal{O}\!\left(L^{-2}\right) 	\nn\\
    \hat{u}_a		&= u_a + \mathcal{O}\!\left(L^{-1}\right)
    \label{eq: limit of chemical potentials}
\end{align}
for any constant $\rho$.

\section{Black holes in the gravitational dual}\label{sec: BHs}

So far we have considered a generic CFT in $d$ dimensions; now let us consider one which admits a dual gravitational description via the AdS/CFT correspondence \cite{Maldacena:1997re}. Generically, this gravity theory exists on spacetimes that are asymptotically of the form AdS$_{d+1}\times \Sigma$, for some compact space $\Sigma$. Then, by taking the DLCQ on both sides of this duality, one arrives at a gravitational dual for the DLCQ of the CFT \cite{Son:2008ye,Goldberger:2008vg,Balasubramanian:2008dm,Herzog:2008wg,Barbon:2008bg,Adams:2008wt,Maldacena:2008wh}.

In more detail, one can write AdS$_{d+1}$ in coordinates that realise a foliation by pp-wave spacetimes,
\begin{align}
  ds^2_{\text{AdS}} = \frac{dr^2}{g^2r^2} + r^2 \Big(-2dx^+ dx^- - x^i x^i (dx^-)^2 + dx^i dx^i\Big) - \frac{1}{g^2}(dx^-)^2
  \label{eq: AdS pp-wave slicing}
\end{align}
with $g=R_\text{AdS}^{-1}$ the inverse AdS radius, and $i=1,\dots,d-2$. In particular, the metric (\ref{eq: AdS pp-wave slicing}) arises precisely in the Penrose limit of global coordinates on AdS \cite{Maldacena:2008wh}, as reviewed below. We denote by $\widetilde{\text{AdS}}_{d+1}$ the spacetime (\ref{eq: AdS pp-wave slicing}) subject to the null identification $x^+\sim x^+ + 2\pi$. Then, $\widetilde{\text{AdS}}_{d+1}$ realises as its conformal boundary ($r\to \infty$) the null-compactified pp-wave spacetime (\ref{eq: pp-wave}). Thus, the DLCQ of the original CFT admits a gravitational dual, formulated on spacetimes asymptotic to $\widetilde{\text{AdS}}_{d+1} \times \Sigma$. The dictionary between parameters on each side of the duality is inherited from that of the original relativistic duality.\\

Next, one would like to construct black hole solutions with $\widetilde{\text{AdS}}_{d+1} \times \Sigma$ asymptotics. We follow an approach first considered in \cite{Maldacena:2008wh}. The basic idea is to begin with a known $\text{AdS}_{d+1} \times \Sigma$ black hole in the dual of the parent CFT$_d$, and take the limit constructed in Section \ref{sec: lens to lc} in the bulk theory.

 For simplicity of notation let us specialise once more to $d=2(m+1)$, although the following holds similarly for $d$ odd. By definition, an asymptotically AdS$_{d+1}$ black hole\footnote{We suppress the geometry in the internal space $\Sigma$.} admits coordinates $(\hat{r},\hat{t},\hat{n}_\mu,\hat{\phi}_\mu)$, $\mu =1,\dots, m+1$, with $\hat{n}_\mu>0$ satisfying $\sum_\mu \hat{n}_\mu^2 = 1$, such that we find the $r\to\infty$ asymptotics
\begin{align}
  ds_\text{BH}^2 \quad \longrightarrow\quad & - \frac{(1+g^2 \hat{r}^2)}{g^2}d\hat{t}^2 + \frac{d\hat{r}^2}{1+g^2 \hat{r}^2} + \hat{r}^2 \sum_{\mu=1}^{m+1} \left(d\hat{n}_\mu^2 + \hat{n}_\mu^2 d\hat{\phi}_\mu^2\right) + \dots 		\nn\\
  & = \frac{d\hat{r}^2}{g^2 \hat{r}^2} + \hat{r}^2 \left(-d\hat{t}^2 + \sum_{\mu=1}^{m+1} \left(d\hat{n}_\mu^2 + \hat{n}_\mu^2 d\hat{\phi}_\mu^2\right) \right) + \dots 
\end{align}
thus realising an $\mathbb{R}_t\times S^{2m+1}$ conformal boundary. The bulk dual of the Penrose rescaling (\ref{eq: Penrose limit coordinate transformation}) is to define new coordinates $(r,x^+, x^-, n_\alpha, \phi_\alpha)$, $\alpha = 1,\dots, m$ by
\begin{align}
  r = \frac{1}{\eta} \hat{r},\quad x^+ = \eta^2 (\hat{t} + \hat{\phi}_{m+1}),\quad x^-  = \hat{t},\quad n_\alpha = \eta \hat{n}_\alpha, \quad \phi_\alpha = \hat{\phi}_\alpha 
  \label{eq: bulk Penrose limit}
\end{align}
and take $\eta \to\infty$. In doing so, we arrive at a solution with the asymptotics (\ref{eq: AdS pp-wave slicing}), with $(n_\alpha,\phi_\alpha)$ providing two sets of polar coordinates in the two planes $\mathbb{R}^4=\mathbb{R}^2\times \mathbb{R}^2$ spanned by the $\{x^i\}$. We can finally arrive at an asymptotically $\widetilde{\text{AdS}}_{d+1}$ solution by identifying the coordinate $x^+\sim x^+ + 2\pi$ throughout the whole spacetime.\\

Any initial black hole solution will depend on a number of parameters $\{b\}$. In taking this Penrose limit, we must also specify the scaling behaviour of such parameters. Indeed, it is clear by the rescaling $r = \eta^{-1}\hat{r}$ that if we simply hold the $\{b\}$ fixed, as we take $\eta \to\infty$ we simply zoom in on the asymptotic geometry, and end up with precisely pure $\widetilde{\text{AdS}}_{d+1}$ as in (\ref{eq: AdS pp-wave slicing}). It was first demonstrated in \cite{Maldacena:2008wh} that this issue can be circumvented, and the black hole retained, for rotating black holes in various dimensions.

Associated to any such black hole in $d+1=2m+3$ dimensions is a mass $\hat{M}$, angular momenta $\{\hat{J}_\mu\}$, and a Bekenstin-Hawking entropy $\hat{\mathcal{S}}$. Amongst the parameters $\{b\}$ one has a subset $\{a_\mu\}$ taking values in $(-g^{-1},g^{-1})$ which are in a sense dual to the angular momenta $\{\hat{J}_\mu\}$; in particular, taking $|ag|\to 1$ corresponds to a limit of large $|J_\mu|$. We may also have some additional quantities like electric charges, but they will not play a key role here.

Then, the Penrose limit is effectively an infinite boost along the $\mu=m+1$ direction. In order to retain the black hole as we move to this strongly boosted frame, we must perform a commensurate limit of $a_\mu$ in order that the black hole ``keeps up'' with the frame. This can be made very concrete at the level of the explicit supergravity solutions \cite{Maldacena:2008wh,Klemm:2014rda,Hennigar:2014cfa,Hennigar:2015cja}. One requires then\footnote{The sign here is ambiguous only because the overall sign of the $\{a_\mu\}$ relative to that of the $\{J_\mu\}$ varies between different black hole solutions in the literature. The sign we want is that one in which $\hat{J}_{m+1}$ becomes large and positive.}
\begin{align}
  ga_{m+1} = \pm \left(1- \frac{1}{2\lambda \eta^2}\right) \quad \longrightarrow \quad \pm 1
  \label{eq: a limit}
\end{align}
for some $\lambda>0$, which becomes a new parameter of the limiting solution\footnote{In \cite{Hennigar:2014cfa,Hennigar:2015cja}, the limit used effectively has $\lambda=1$. But this additional degree of freedom is re-introduced by imposing the periodicity $x^+ \sim x^+ + \mu$ for a generic $\mu$. It is straightforward to show that $\lambda$ and $\mu$ parameterise effectively the same degree of freedom in the solution, by rescaling the coordinates $\{x^+, x^-, x^i\}$ appropriately.}. After imposing the identification $x^+ \sim x^+ + 2\pi$, we arrive at an asymptotically $\widetilde{\text{AdS}}_{d+1}$ black hole. We can compute from this solution a mass $\Delta$, angular momenta $\{J_\alpha\}$ corresponding to rotations transverse to the chosen lightcone, a momentum $K$ along the $x^+$ direction, and finally a Bekenstein-Hawking entropy $\mathcal{S}=A/4G_N$. All of these quantities (as well as any additional charges) can be equivalently found as a limit of those of the parent black hole, as\footnote{The additional scaling by $\eta^{-2}$ relative to (\ref{eq: field theory charge rescaling}) arises due to the identification $x^+ \sim x^+ + 2\pi$, which can be formulated at finite $\eta$ as an identification $(\hat{r},\hat{t},\hat{n}_\mu,\hat{\phi}_\mu)\sim (\hat{r},\hat{t},\hat{n}_\mu,\hat{\phi}_\mu+ \frac{2\pi}{L}) $ for $L = \eta^2 \in \mathbb{N}$. This has the effect of quotienting the spacetime, and in particular the horizon, by a (free) $\mathbb{Z}_L$ action. This induces the stated rescaling of quantities.}
\begin{align}
  \Delta = \lim_{\eta\to\infty} \frac{1}{\eta^2}\!\left(\hat{M}-\hat{J}_{m+1}\right),\quad J_\alpha = \lim_{\eta\to\infty} \frac{1}{\eta^2} \hat{J}_\alpha,\quad K = \lim_{\eta\to\infty} \frac{1}{\eta^4} \hat{J}_{m+1},\quad \mathcal{S} = \lim_{L\to\infty} \frac{1}{\eta^2}\hat{\mathcal{S}}
\end{align}
where it is implicit that we are also taking the behaviour (\ref{eq: a limit}) for $a_{m+1}$ as we take the limit, while we are free to keep all other parameters fixed. Indeed, it is precisely this behaviour which ensures that the quantities $\Delta, J_\alpha, K$ and $\mathcal{S}$ are generically non-zero, and thus in particular that the horizon area is finite and non-zero.

Precisely this form of limit was first considered in \cite{Maldacena:2008wh} for non-supersymmetric rotating black holes in AdS$_5$ and AdS$_7$. An important observation was made there, that in order to regard such black holes consistently as solutions in the supergravity approximation to string/M-theory, one requires the compact direction to become large and spacelike in the bulk. This is in turn achieved provided the momentum $K$ is sufficiently large; exactly what regime one requires $K$ to be in is model dependent.\\[-0.7em]

Finally, let us briefly review the existing literature on limits of rotating black holes of the form (\ref{eq: a limit}), which are generally referred to as \textit{ultra-spinning black holes}. This regime of parameter space was first considered in \cite{Emparan:2003sy} for asymptotically flat Myers-Perry back holes \cite{Myers:1986un}, and then later in \cite{Caldarelli:2008pz,Caldarelli:2011idw} for Kerr-Newman-AdS black holes, as are relevant to our discussion here. The key observation is that as $|ag|\to\pm 1$ for some rotation parameter $a$, the Kerr-Newman-AdS metric diverges, and so one must perform a compensating scaling limit of coordinates to arrive at a well-defined solution. Different choices of such a coordinate limit give rise to distinct solutions. Although these early works did identify such a coordinate rescaling, the Penrose rescaling (\ref{eq: bulk Penrose limit}) was not considered until \cite{Maldacena:2008wh}, where ultra-spinning black holes of the form we are interested in were first derived in AdS$_5$ and AdS$_7$. For the remainder of this paper, we use ``ultra-spinning'' to denote specifically those black holes found by a simultaneous $|ag|\to\pm 1$ and Penrose limit.

A little later, ultra-spinning solutions in AdS$_4$ were studied briefly in \cite{Gnecchi:2013mja} and in more detail in \cite{Klemm:2014rda}; this latter work in particular recovered the solutions of \cite{Caldarelli:2008pz,Caldarelli:2011idw} by zooming into a particular near-horizon region of the ultra-spinning black hole geometry. A broader analysis of ultra-spinning black holes was given shortly afterwards in \cite{Hennigar:2014cfa,Hennigar:2015cja}, who formulated the ultra-spinning limit of Kerr-AdS solutions in general dimensions. Of particular relevance to the present work is the derivation in \cite{Hennigar:2015cja} of the ultra-spinning limit of the AdS$_5$ black hole of \cite{Chong:2005hr} in a five-dimensional supergravity theory, which is dual to an ensemble of states in $\N=4$ super-Yang-Mills. More recently, the ultra-spinning limit of the AdS$_6$ black hole of \cite{Chow:2008ip} in six-dimensional gauged supergravity was considered in \cite{Wu:2021vch}. See also \cite{Wu:2020cgf,Wu:2020mby} for other recent developments.

Lastly, in \cite{Dorey:2022cfn} the present authors considered the ultra-spinning black holes present in the dual of the DLCQ of the  six-dimensional $(2,0)$ theory. Such solutions can be constructed in the ultra-spinning limit of AdS$_7$ black holes in gauged supergravity in seven dimensions. The solution found in this work arose as the ultra-spinning limit of the solution of \cite{Chow:2007ts}. The details of this derivation will appear elsewhere. 

 
\newpage 
{\noindent \LARGE\bf\color{mygreen} Part II: Examples, and their gravity duals}	
\addcontentsline{toc}{section}{\color{mygreen}\large Part II: Examples, and their gravity duals} 

\section{The six-dimensional $(2,0)$ theory}\label{sec: 6d}

Let us now study the example of the six-dimensional non-Abelian $(2,0)$ superconformal field theory. This is a non-Abelian gauge theory for some simply-laced gauge group, possibly with some Abelian factors too; we specialise to the case of $U(N)$. Then, the $U(N)$ $(2,0)$ theory is the model living on a stack of $N$ M5-branes in M-theory. In the language of the previous Sections, we have $d=4$ and $m=2$. 

The symmetry algebra of the theory is the Lie superalgebra $\frak{osp}(8^*|4)$. As a six-dimensional conformal field theory, the spacetime symmetries form $SO(2,6)$. The R-symmetries make up $SO(5)$, which correspond to rotations in the directions transverse to the M5-branes; we choose a pair of Cartan generators $Q_a$, $a=1,2$ for this $SO(5)$. The theory has 16 real Poincar\'e supercharges (i.e. $\Q$'s), and a further 16 real conformal supercharges (i.e. $\S$'s). Finally, let us denote by $n_I$, $I=1,\dots,N$ the Cartan generators for the global part of the $U(N)$ gauge group, under which gauge non-invariant states are charged.\\

We now give an overview of our results in this setting. A key feature of this model that we will exploit is that its DLCQ description is well-understood \cite{Aharony:1997th,Aharony:1997an}. Namely, the DLCQ theory at some fixed particle number $K$ precisely coincides with the quantum mechanics of $K$ D0-branes in the worldvolume of $N$ coincident D4-branes; in the language of field theory, this is the quantum mechanics describing the slow motion of $K$ $\frac{1}{2}$-BPS instanton-particles in $\N=2$ super-Yang-Mills in five dimensions. It follows that the lightcone partition function $\Z$ can be decomposed into a sum over the partition functions $\Z^{(K)}$ of these quantum mechanical models. We therefore have two independent means by which to compute $\Z$: either by computing the quantum mechanical partition functions $\Z^{(K)}$, or alternatively by taking the appropriate limit of the lens space partition function $\hZ_L$ as set out in Section \ref{sec: lens to lc}.

Neither of these routes is tractable for general values of the chemical potentials of $\Z$. To proceed, we will specialise to a codimension-1 subspace of chemical potential space on which $\Z$ becomes a superconformal index, $\Z\to \I$. We will then first compute $\I$ as a sum over the superconformal indices $\I^{(K)}$ of each theory of instanton quantum mechanics. Conversely, the results of Section \ref{sec: lens to lc} imply that $\I$ can be found in a an appropriate limit of the lens space superconformal index $\hI_L$ \cite{Kim:2013nva}. We take this limit, and verify that it matches precisely the result from quantum mechanics.

Finally, we will briefly review the work \cite{Dorey:2022cfn}, which studies the holographic dual of this instanton quantum mechanics, and in particular the ultra-spinning black hole solutions it admits.

\subsection{DLCQ as instanton quantum mechanics}

Let us describe in more detail the DLCQ of the 6d $U(N)$ $(2,0)$ theory. Upon null compactification the spacetime symmetry $SO(2,6)$ is broken to the Schr\"odinger group $\text{Schr}(4)$, while it is clear that the $SO(5)$ R-symmetry is unaffected. Finally, of the $32$ real Poincar\'e and conformal supercharges of $\frak{osp}(8^*|4)$, only $24$ are preserved \cite{Aharony:1997an}\footnote{See also \cite{Sakaguchi:2008rx}.} . In lieu of a common notation in the literature, we denote this sub-superalgebra of $\frak{osp}(8^*|4)$ by\footnote{One can also consider the DLCQ of $(1,0)$ superconformal theories in six-dimensions, which have an $SU(2)$ R-symmetry and $8+8=16$ real supercharges; in this case, the resulting superalgebra has $SU(2)$ R-symmetry and 12 real supercharges. In this paper however we always use super-Schr$(4)$ to denote the maximally-supersymmetric case found in the DLCQ of the $(2,0)$ superconformal algebra.} super-$\frak{Schr}(4)$, with corresponding group super-Schr(4). The lightcone partition function (\ref{eq: lc pf}) is defined as a trace over gauge-invariant states.

The sector of this theory of fixed null momentum $K\ge 0$ is identified precisely as a non-linear $\sigma$-model in $(0+1)$-dimensions, whose target space is $\mathcal{M}_{K,N}$, the moduli space of $K$ instantons in a $U(N)$ Yang-Mills theory. A detailed analysis of how one properly defines such a model, and subsequently computes its superconformal index, can be found in \cite{Dorey:2022cfn}. Here, we just review the essential facts.

The fact that $\M_{K,N}$ is a hyper-K\"ahler cone ensures that the corresponding $\sigma$-model is a theory of maximally supersymmetric superconformal quantum mechanics, with superalgebra $\frak{osp}(4^*|4)$ and $8+8$ real supercharges. There are some further global symmetries specific to these models. There is firstly an $SU(2)$ global symmetry. Then, $\frak{osp}(4^*|4)\oplus \frak{su}(2)\subset \text{super-}\frak{Schr}(4)$, where the $\frak{sl}(2,\mathbb{R})$ subalgebra of $\text{super-}\frak{Schr}(4)$ generated by $\{H,C,D\}$ is precisely the $(0+1)$-dimensional conformal algebra $\frak{so}(2,1)\subset \frak{osp}(4^*|4)$. However there are elements of $\text{super-}\frak{Schr}(4)$, such as 8 of the 24 supercharges, which do not appear in $\frak{osp}(4^*|4)\oplus \frak{su}(2)$. One can show however that all such elements are in fact realised (albeit non-linearly) in the quantum mechanics on $\M_{K,N}$ \cite{Aharony:1997an}, as is required for the DLCQ interpretation to make sense. In summary, we have the chain of inclusions
\begin{align}
  \frak{osp}(4^*|4)\oplus \frak{su}(2) \subset \text{super-}\frak{Schr}(4) \subset \frak{osp}(8^*|4)
\end{align}
none of which are strict.

Finally, the quantum mechanics also has an $SU(N)$ global symmetry, which corresponds to (the simple part of) the global part of the gauge group $U(N)$ in the six-dimensional theory. Its generators are of the form $n_I - n_J$ for $I\neq J$.

Let us then define the \textit{singlet instanton partition function}
\begin{align}
  \Z^{(K)}(\mu,w_\alpha,u_a) = \text{Tr}^\text{singlets}_{\mathcal{M}_{K,N}}\exp \Big[ -\mu \Delta - w_\alpha J_\alpha - u_a Q_a\Big]
\end{align}
where the trace is performed over all states on $\M_{K,N}$ that are singlets under the $SU(N)$ global symmetry. The proposal of \cite{Aharony:1997th,Aharony:1997an} implies then that we have precisely
\begin{align}
  \Z(\beta,\mu,w_\alpha,u_a) = \sum_{K=0}^\infty e^{-\beta K} \Z^{(K)}(\mu,w_\alpha,u_a)
  \label{eq: Z QM decomp}
\end{align}
where $\Z^{(0)}=1$.

\subsection{Specialising to the superconformal index}

To make progress to actually compute $\Z$, we specialise to a codimension-1 subspace of chemical potentials on which it becomes an index. We fix
\begin{align}
  2\mu - w_1 - w_2 + u_1 + u_2 = 2\pi i \quad (\text{mod } 4\pi i)
  \label{eq: QM chemical potential constraint}
\end{align}
Then, there exists a supercharge $\Q\in\frak{osp}(4^*|4)$ with conjugate $\S=\Q^\dagger$, such that we have the vanishing anti-commutator
\begin{align}
  \left\{ e^{-\beta K - \mu \Delta - w_\alpha J_\alpha - u_a Q_a}, \Q \right\}  = 0
\end{align}
This is then precisely the statement that on this subspace, $\Z$ is a Witten index for the supercharge $\Q$; for this reason, let us write $\I(\beta,\mu,w_\alpha,u_a)=\Z(\beta,\mu,w_\alpha,u_a)$ for the restriction of $\Z$ to the subspace (\ref{eq: QM chemical potential constraint}). $\I$ then only counts states annihilated by $\Q$ and $\S$. Such states saturate the BPS bound,
\begin{align}
  \{\Q,\S\} = \H := \Delta - J_1 - J_2 - 2Q_1 - 2 Q_2
\end{align}
While naively $\I$ appears to depend on five independent variables, the fact that only states with $\H=0$ contribute to it implies a continuous shift symmetry, given by
\begin{align}
  \I(\beta,\mu,w_\alpha,u_a) = \I(\beta,\mu-\alpha,w_\alpha+\alpha,u_a+2\alpha)
  \label{eq: shift symmetry}
\end{align}
which clearly preserves the constraint (\ref{eq: QM chemical potential constraint}). Thus, $\I$ in effect depends only on four independent variables. One can make this manifest by choosing a maximal set of shift-invariant linear combinations of $\{\beta,\mu,w_\alpha, u_a\}$. A convenient choice is $\{\beta,\epsilon_1,\epsilon_2,m\}$, with
\begin{align}
  \epsilon_1 =  w_1+ \mu,\quad \epsilon_2 = w_2 + \mu,\quad m = u_1  - \tfrac{1}{2}(w_1 + w_2) + \mu
  \label{eq: QM reduced variables}
\end{align}
We refer to these as \textit{reduced} variables. In terms of these reduced variables, we have
\begin{align}
  &\I(\beta,\mu,w_\alpha,u_a) = \text{Tr}_\text{LC}(-1)^F e^{-\beta K - \epsilon_1 \left(J_1+\frac{1}{2}(Q_1 + Q_2)\right) - \epsilon_2\left(J_2 + \frac{1}{2}(Q_1+Q_2)\right) - m(Q_1 - Q_2)} 
  \label{eq: 6d LC index def}
\end{align}

\subsection{Result from quantum mechanics}\label{subsec: 6d result from QM}

Following (\ref{eq: Z QM decomp}), we have
\begin{align}
  \I(\beta,\mu,w_\alpha,u_a) = \sum_{K=0}^\infty e^{-\beta K} \I^{(K)}(\mu,w_\alpha,u_a)
\end{align}
where $\I^{(K)}$ is just $\Z^{(K)}$ constrained to the constraint surface (\ref{eq: QM chemical potential constraint}), and hence
\begin{align}
  &\I^{(K)}(\mu,w_\alpha,u_a)= \text{Tr}^\text{singlets}_{\mathcal{M}_{K,N}}(-1)^F e^{ - \epsilon_1 \left(J_1+\frac{1}{2}(Q_1 + Q_2)\right) - \epsilon_2\left(J_2 + \frac{1}{2}(Q_1+Q_2)\right) - m(Q_1 - Q_2)} 
\end{align}
As an index, $\I^{(K)}$ is invariant under continuous deformations of the theory, which in this context simply means continuous deformations of the $\sigma$-model target space\footnote{Or more correctly, the resolved space $\tilde{\M}_{K,N}$ on which one should formulate the quantum mechanics; see \cite{Dorey:2022cfn} and references therein.} $\M_{K,N}$. In other words, it is a topological invariant, identified as a particular Euler characteristic for equivariant sheaf cohomology \cite{Barns-Graham:2018xdd}. This can in turn be computed by localisation theorems in equivariant K-theory, from which one finds the result
\begin{align}
  \I^{(K)}{(\mu,w_\alpha,u_a)} = \oint d\mu_\text{Haar}[\lambda] \Z_\text{inst}^{(K)}(\epsilon_1,\epsilon_2,m,\lambda_I)
\end{align}
in terms of Haar measure
\begin{align}
  \oint d\mu_\text{Haar}[\lambda] = \left(\prod_{I=1}^N \oint_0^{2\pi} \frac{d\lambda_I}{2\pi} \right) \prod_{I<J} \left(2\sin \frac{\lambda_{IJ}}{2}\right)
  \label{eq: Haar measure}
\end{align}
Here, $\Z^{(K)}_\text{inst}$ is the $K$-instanton contribution to the non-perturbative part of the Nekrasov partition function of an auxiliary $\N=1^*$ $SU(N)$ five-dimensional super-Yang-Mills theory, with $\Omega$-deformation parameters $\epsilon_1,\epsilon_2$, adjoint hypermultiplet mass $m$, and Coulomb branch parameters $z_I = e^{-i \lambda_I}$, where recall $I=1,\dots,N$. The integral against the Haar measure projects us onto $SU(N)$-invariant states as required, where the $\lambda_I$ contours run from $0$ to $2\pi$ along the real line assuming that we have $\text{Re}(\epsilon_{1,2})>0$. Further details on $\Z_\text{inst}^{(K)}$ including its explicit form can be found in Appendix \ref{app: 6d conventions}.

The total non-perturbative contribution to this Nekrasov partition function is
\begin{align}
  \Z_\text{inst}(\beta,\epsilon_1,\epsilon_2,m,\lambda_I) = \sum_{K=0}^\infty e^{-\beta K} \Z_\text{inst}^{(K)}(\epsilon_1,\epsilon_2,m,\lambda_I)
  \label{eq: total non-pert Nek}
\end{align}
in terms of which the lightcone superconformal index is simply
\begin{align}
  \I(\beta,\mu,w_\alpha,u_a) = \oint d\mu_\text{Haar}[\lambda]  \Z_\text{inst}(\beta,\epsilon_1,\epsilon_2,m,\lambda_I)
  \label{eq: LC index QM result}
\end{align}
Let us briefly make contact with another avatar of $\Z_{\text{inst}}$. As we will review in detail in Section \ref{sec: 4d}, a wide variety of supersymmetric partition functions and indices of theories in three and four dimensions admit a form as a product of three-dimensional holomorphic blocks, suitably glued together, and integrated over to project onto gauge-invariant states. It turns out that a similar phenomenon holds for certain quantities in five \cite{Nieri:2013vba,Pasquetti:2016dyl} and six \cite{Kim:2013nva,Kim:2016usy} dimensions. In this setting, the basic building block is the so-called five-dimensional holomorphic block $\mathcal{B}$, which can be understood as the Nekrasov partition function of a relevant five-dimensional $\N=1$ theory on an Omega background, $S^1\times \mathbb{R}^4_{\epsilon_1,\epsilon_2}$. Schematically, such blocks take the form
\begin{align}
  \mathcal{B} = \Z_\text{pert}\Z_{\text{inst}}
\end{align}
where $\Z_{\text{pert}}$ is the perturbative contribution, which in the localisation computation arises as a product of classical and one-loop contributions. Indeed, the form of $\Z_{\text{pert}}$ is somewhat ambiguous, as different choices can give rise to the same partition functions upon glueing \cite{Nieri:2013vba}. Then, the superconformal index of the DLCQ theory (\ref{eq: LC index QM result}) coincides precisely with the non-perturbative part of the holomorphic block of the $\N=1^*$ theory mentioned above, projected onto gauge singlets \cite{KimLee}.

\subsection{Result from the lens space index}

Let us now compute the lightcone index $\I(\beta, \mu,w_\alpha,u_a)$ once again as a limit of the index of the theory on lens space, following Section \ref{sec: lens to lc}. 

\subsubsection{An orbifold and an index}

We now consider the theory on $\mathbb{R}_t\times S^5/\mathbb{Z}_L[p_\alpha^L]$ found by orbifolding the 6d $(2,0)$ theory by the action of
\begin{align}
  g_L = \exp\left(\frac{2\pi i}{L}\left(p^L_1 \hat{J}_1 + p_2^L \hat{J}_2 + \hat{J}_3 + \alpha_1 Q_1 + \alpha_2 Q_2\right)\right)
  \label{eq: gL 6d}
\end{align}
and consider the refined partition function $\hZ_L(\hat{\mu},\hat{w}_\mu,\hat{u}_a)$ as defined in (\ref{eq: lens pf}). We next need to specialise both our choice of orbifold parameters $(p_1^L,p_2^L,\alpha_1,\alpha_2)$ and chemical potentials $(\hat{\mu},\hat{w}_\mu,\hat{u}_a)$ such that the following conditions are satisfied:
\begin{enumerate}
  \item The orbifold preserves the supercharge $\Q$ with respect to which the lightcone superconformal index $\I$ is an index
  \item The orbifold is \textit{good}, so that we don't quotient by any internal symmetries 

  \item $\hZ_L$ becomes a superconformal index $\hat{\Z}_L\to \hI_L$ for the supercharges $\{\Q,\S=\Q^\dagger\}$
  \end{enumerate}
The 16 real Poincar\'e supercharges of $\frak{osp}(8^*|4)$ carry charges $(\hat{J}_1,\hat{J}_2,\hat{J}_3,Q_1,Q_2)= (\pm 1)^5$, with the product of the three angular momenta constrained to be $(-1)$. Then, the supercharge $\Q$ turns out to be the one with charges $(-,-,-,+,+)$. Condition 1 then requires $g_L \Q=\Q$ for all $L$. It is clear this is satisfied if we take $p_1^L=p_2^L=1$, so that the orbifold acts precisely along the Hopf fibre of $S^5$, and
\begin{align}
  \alpha_1 = \frac{3}{2}+\hat{n},\quad \alpha_2 = \frac{3}{2} - \hat{n}
\end{align}
for some $\hat{n}\in \mathbb{R}$. Next, it is argued in \cite{Kim:2013nva} that Condition 2 is satisfied provided we take $\hat{n}\in \frac{2\mathbb{Z}+1}{2}$. In particular, the special choices of $\hat{n}=\pm \frac{1}{2},\pm\frac{3}{2}$ preserve additional supercharges.

Finally, Condition 3 is met by fixing
\begin{align}
  \hat{\mu}-\hat{w}_1 - \hat{w}_2 - \hat{w}_3 + \hat{u}_1 + \hat{u}_2 =2\pi i \quad (\text{mod } 4\pi i)
  \label{eq: 6d chemical potential constraint}
\end{align}
We write $\hI_L(\hat{\mu},\hat{w}_\mu,\hat{u}_a)=\hZ_L(\hat{\mu},\hat{w}_\mu,\hat{u}_a)$ when restricted to this subspace. Then, $\hI_L$ counts only those states that are annihilated by $\Q,\S$. Such states saturate the BPS bound
\begin{align}
  \{\Q,\S\} =\H = \hat{H} - \hat{J}_1 - \hat{J}_2 - \hat{J}_3 - 2Q_1 - 2Q_2 \ge 0
\end{align}
While naively $\hI_L$ appears to depend on five independent variables, the fact that only states with $\H=0$ contribute to it implies a continuous shift symmetry, given by
\begin{align}
  \hI_L(\hat{\mu},\hat{w}_\mu,\hat{u}_a) = \hI_L(\hat{\mu}-\alpha,\hat{w}_\mu+\alpha,\hat{u}_a+2\alpha)
  \label{eq: shift symmetry 6d}
\end{align}
which clearly preserves the constraint (\ref{eq: QM chemical potential constraint}). Thus, $\hI_L$ only actually depends on four independent variables. Much like with the lightcone index, we can make this manifest by choosing \textit{reduced} variables: a maximal set of shift-invariant linear combinations of $\{\hat{\mu},\hat{w}_\mu,\hat{u}_a\}$. A convenient choice is $\{\hat{\beta},\hat{v}_1,\hat{v}_2,\hat{m}\}$, defined by
\begin{align}
  \hat{\beta} = \hat{\mu}+\tfrac{1}{3}(\hat{w}_1+\hat{w}_2+\hat{w}_3),\quad \hat{v}_\mu = \hat{w}_\mu - \tfrac{1}{3}(\hat{w}_1+\hat{w}_2+\hat{w}_3),\quad \hat{m} = \hat{u}_1 + \tfrac{1}{2}(\hat{\mu} - \hat{w}_1-\hat{w}_2-\hat{w}_3)
  \label{eq: 6d reduced variables}
\end{align}
where we have also defined $\hat{v}_3=-\hat{v}_1 - \hat{v}_2$ for the sake of symmetry in later expressions. In terms of these reduced variables, $\hI_L$ takes the form
\begin{align}
  \hI_L(\hat{\mu},\hat{w}_\mu,\hat{u}_a) = \text{Tr}\,(-1)^F\exp \Big[-\hat{\beta}\!\left(\hat{H} - \tfrac{1}{2}(Q_1 + Q_2)\right) - \hat{v}_\mu \hat{J}_\mu - \hat{m}(Q_1-Q_2)\Big]
  \label{eq: 6d lens index}
\end{align}

\subsubsection{Explicit form of the six-dimensional index}

A series of papers \cite{Kim:2012qf,Kim:2013nva} (see also the review \cite{Kim:2016usy} and references therein) tackles precisely the issue of computing $\hI_L$. A method is proposed\footnote{The are a number of assumptions that enter into this proposal. For instance, for $L>1$ one should in principle worry about summing over discrete gauge holonomies, as discussed in Section \ref{sec: lens space theory}. It is argued in \cite{Kim:2013nva} however that in this case, such sectors are not present. One can view our results as a confirmation of the validity of this and others assumptions, at least in the $L\to\infty$ limit.} which exploits the local decomposition $\mathbb{R}\times S^5/\mathbb{Z}[1,1] \sim \mathbb{R}\times S^1 \times \mathbb{CP}^2 $ to recast $\hI_L$ as a path integral of a certain five-dimensional supersymmetric gauge theory on $\mathbb{R}\times \mathbb{CP}^2$ \cite{Kim:2012tr}. This path integral is then subsequently computed exactly via supersymmetric localisation.

Let us here simply state the result of this work. We specialise to the case of $\hat{n}=-\frac{1}{2}$, for which the resulting index takes a simpler form. Then, one finds
\begin{align}
  &\I_L(\hat{\mu},\hat{w}_\mu,\hat{u}_a) \nn\\
  &= \sum_{s_1,\dots,s_N=-\infty}^\infty \oint \prod_{I=1}^N\left(\frac{d\lambda_I}{2\pi}\right) e^{-LS(\lambda,s,\hat{\beta})}\Z^{(s)}_\text{pert}(\hat{\beta}, \hat{v}_\alpha,\hat{m};\lambda)\Z^{(s)}_\text{non-pert}(\hat{\beta}, \hat{v}_\alpha,\hat{m};\lambda,L)
  \label{eq: IL explicit result}
\end{align}
Note that the integrand manifestly depends on the chemical potentials $\{\hat{\mu},\hat{w}_\mu,\hat{u}_a\}$ only through the reduced variables $\{\hat{\beta},\hat{v}_\alpha,\hat{m}\}$ as defined in (\ref{eq: 6d reduced variables}). As well as these variables, the integrand depends on $N$ complex fugacities $\lambda_I\sim \lambda_I + 2\pi$ for the Cartan generators of $\frak{u}(N)$, and $N$ integers $s_I$ which arise in the computation from instantons on $\mathbb{CP}^2$ carrying self-dual flux. The non-perturbative contribution $\Z_\text{non-pert}$ is constructed from $\Z_\text{inst}$ as defined in (\ref{eq: total non-pert Nek}), as 
\begin{align}
 \Z^{(s)}_\text{non-pert}(\hat{\beta}, \hat{v}_\alpha,\hat{m};\lambda,L) = \prod_{\mu=1}^3 \Z_\text{inst}\left(\beta^{(\mu)},\epsilon_1^{(\mu)},\epsilon_2^{(\mu)},m^{(\mu)},\lambda_I^{(\mu)}\right) 
\end{align}
The arguments of each factor are given by
\begin{align}
  \beta^{(\mu)}		&=	L(\hat{\beta}+\hat{v}_\mu)			\nn\\
  \epsilon_1^{(\mu)}	&= 	-\tfrac{3}{2}\hat{v}_\mu	 + \tfrac{1}{4}\epsilon_{\mu\nu\rho}\hat{v}_{\nu\rho} 		\nn\\
  \epsilon_2^{(\mu)}	&=	-\tfrac{3}{2}\hat{v}_\mu	 - \tfrac{1}{4}\epsilon_{\mu\nu\rho}\hat{v}_{\nu\rho} 		 	\nn\\
  m^{(\mu)}			&=	 \hat{m} + \tfrac{1}{2}(\hat{\beta}+\hat{v}_\mu) 			\nn\\
  \lambda_I^{(\mu)}	&= \lambda_I + i \hat{v}_\mu s_I
  \label{eq: Nek parameters of each factor}
\end{align}
where $\hat{v}_{\mu\nu} = \hat{v}_\mu - \hat{v}_\nu$.

The perturbative contribution $\Z_{\text{pert}}$ takes a quite complicated form\footnote{The factorisation of the integrand extends to the perturbative piece: $\Z_\text{pert}$ can be naively written as a product of three factors, each of which is the perturbative contribution to a Nekrasov partition function with the parameters (\ref{eq: Nek parameters of each factor}). However, as pointed out in \cite{Kim:2013nva}, one must correct this picture slightly due to a contribution from ghost multiplets on the lens space.} for generic fluxes $s$; for our purposes, we need only note its form for zero flux, given by
\begin{align}
  \Z^{(s=(0,\dots,0))}_\text{pert}(\hat{\beta}, \hat{v}_\alpha,\hat{m};\lambda) = \prod_{I<J}\left(2\sin \frac{\lambda_{IJ}}{2}\right)^2
\end{align}
for $\lambda_{IJ}=\lambda_I-\lambda_J$, and so in particular we identify 
\begin{align}
   \oint\prod_{I=1}^N \left(\frac{d\lambda_I}{2\pi}\right)\Z^{(s=(0,\dots,0))}_\text{pert}(\hat{\beta}, \hat{v}_\alpha,\hat{m};\lambda) \Big( \dots  \Big) = \oint d\mu_\text{Haar}[\lambda]\Big( \dots  \Big)
  \label{eq: perturbative zero flux}
\end{align}
in terms of the Haar measure (\ref{eq: Haar measure}) on $U(N)$.

Finally, the `classical action' $S(\lambda, s,\hat{\beta})$ is given by
\begin{align}
  S(\lambda, s, \hat{\beta}) =\sum_{I=1}^N \left(-\frac{1}{2}\hat{\beta}s_I^2 + i s_I \lambda_I\right)
  \label{eq: classical action}
\end{align}
Each integral $\int d\lambda_I$ is over a contour that wraps the complex cylinder $\lambda_I\sim \lambda_I + 2\pi$ precisely once in a right-handed sense. The integrand is a periodic meromorphic function of the $\lambda_I$, and hence the value of the integral does not depend on the local properties of the contour, but only where it sits relative to the integrand's poles. We take a prescription for choosing these contours as first set out in \cite{Kim:2013nva}, which is discussed in detail in Appendix \ref{app: 6d limit}.

\subsubsection{Taking the lightcone limit}

Let us now take the lightcone limit to find $\I$ from $\hI_L$. We need to take $L\to\infty$, while at the same time rescaling chemical potentials as in (\ref{eq: limit of chemical potentials}). Furthermore, this limit of chemical potentials must be compatible with the conditions (\ref{eq: 6d chemical potential constraint}) and (\ref{eq: QM chemical potential constraint}). A simple way to do so is to set
\begin{align}
  \hat{\mu} 		&= \mu + \frac{\beta}{2L} ,\nn\\
   \hat{w}_\alpha 	&= w_\alpha + \frac{\gamma\beta}{2L} ,\nn\\
    \hat{w}_{3} 	&= -\mu + \frac{\beta}{2L}  	\nn\\
    \hat{u}_a		&= u_a + \frac{\gamma\beta }{2L}
\end{align}
corresponding to the choice of $\rho=\beta/2$ in (\ref{eq: limit of chemical potentials}), and for any fixed $\gamma\in \mathbb{R}$. We are thus able to consider the limit $L\to\infty$ while remaining on the supersymmetric subspace (\ref{eq: 6d chemical potential constraint}).
The lightcone index is then found as
\begin{align}
  \I(\beta,\mu,w_\alpha, u_a) = \lim_{L\to\infty} \hI_L \!\left(\hat{\mu}=\mu+\tfrac{\beta}{2L},\,\hat{w}_\alpha=w_\alpha+ \tfrac{\gamma\beta}{L},\,\hat{w}_3=-\mu+\tfrac{\beta}{2L},\,\hat{u}_a=u_a+ \tfrac{\gamma\beta}{L}\right)
  \label{eq: I from IL limit}
\end{align}
Recall, the lens space index $\hI_L$ is expressible as a function only of the four reduced variables $\{\hat{\beta},\hat{v}_\alpha,\hat{m}\}$, while the lightcone index is expressible as a function only of the four reduced variables $\{\beta,\epsilon_\alpha,m\}$. In terms of these, the expression (\ref{eq: I from IL limit}) is recast as
\begin{align}
  \I(\beta,\epsilon_\alpha,m) = \lim_{L\to\infty} \hI_L\!\left(\hat{\beta}=\tfrac{2}{3}\epsilon_+ + \tfrac{(2+\gamma)\beta}{3L},\,\hat{v}_\alpha=\epsilon_\alpha - \tfrac{2}{3}\epsilon_+ - \tfrac{(1-\gamma)\beta}{6L},\,\hat{m}=m\right)
\end{align}
where $\epsilon_\pm = \frac{1}{2}(\epsilon_1\pm \epsilon_2)$. A particularly simple choice for the free parameter is $\gamma=1$, for which we have
\begin{align}
  \I(\beta,\epsilon_\alpha,m) = \lim_{L\to\infty} \hI_L\!\left(\hat{\beta}=\tfrac{2}{3}\epsilon_+ + \tfrac{\beta}{L},\,\hat{v}_\alpha=\epsilon_\alpha - \tfrac{2}{3}\epsilon_+,\,\hat{m}=m\right)
\end{align}
We are then able to take this limit explicitly, at least in the subspace of parameter space defined by
\begin{align}
  \text{Re}(\epsilon_{1,2}) >0
  \label{eq: nice subspace}
\end{align}
The full details of this computation can be found in Appendix \ref{app: 6d limit}. For now, let us just give an outline. The calculation proceeds in two steps:
\begin{enumerate}
  \item Making use of a judicious contour deformation, one can ensure that the $e^{-LS}$ part of the integrand provides an exponential suppression at large $L$ of all terms in $\hI_L$ with flux $s\neq (0,\dots,0)$. Thus, all non-zero flux terms vanish identically as we take $L\to\infty$.
  \item It remains then to study the zero flux term. It is firstly straightforward to see that the first and second factors of $\Z_\text{inst}$ appearing in $\Z_\text{non-pert}$ simply approach $1$ as we take the limit, since $\beta^{(\alpha)} = L(\hat{\beta}+\hat{v}_\alpha) = L\epsilon_\alpha + \beta$ provides a exponential suppression $e^{-Lk\epsilon_\alpha}$ to the $k^\text{th}$ term in the sum over instanton sectors by virtue of the condition (\ref{eq: nice subspace}). In contrast, since $\beta^{(3)}=L(\hat{\beta}+\hat{v}_3) = \beta=\mathcal{O}(1)$, there is no such suppression of the third factor, which therefore remains. In particular, in the limit we find
  \begin{align}
  \lim_{L\to\infty}(\beta^{(3)},\epsilon_1^{(3)},\epsilon_2^{(3)},m^{(3)}) = \lim_{L\to\infty} \left(\beta,\epsilon_1,\epsilon_2,m+\tfrac{\beta}{2L}\right) = \left(\beta,\epsilon_1,\epsilon_2,m\right)
\end{align}
while clearly $\lambda_I^{(3)} = \lambda_I$ as we are in the zero flux sector.
\end{enumerate}
Further using the expression (\ref{eq: perturbative zero flux}) for the zero-flux perturbative contribution, we finally arrive at
\begin{align}
  \I(\beta,\mu,w_\alpha,u_a)& = \int d\mu_{\text{Haar}}[\lambda]\, \Z_{\text{inst}}(\beta,\epsilon_1, \epsilon_2,m, \lambda_I)
  \label{eq: 6d final field theory index}
\end{align}
matching exactly the form of the result (\ref{eq: LC index QM result}) computed directly in instanton  quantum mechanics. Finally, to ensure a precise match, we must be sure that the $\lambda_I$ contours are equivalent to those of (\ref{eq: LC index QM result}). This is rather non-trivial, since the contours in (\ref{eq: 6d final field theory index}) are not free to choose, but instead are entirely fixed by the contour prescription of the lens space index we started with \cite{Kim:2013nva}. Happily, as shown in Appendix \ref{app: 6d limit}, the $\lambda_I$ contour we find in the limit of the lens space are equivalent to the simple straight line connecting $0$ to $2\pi$ along the real line. Thus, the two results (\ref{eq: LC index QM result}) and (\ref{eq: 6d final field theory index}) for the lightcone index match precisely.


\subsection{Ultra-spinning black holes}

The six-dimensional $U(N)$ $(2,0)$ theory admits a gravitational dual description, as M-theory on asymptotically AdS$_7\times S^4 $ spacetime with $N$ units of four-form flux on $S^4$ \cite{Maldacena:1997re}. Following the discussion in Section \ref{sec: BHs}, we find that the superconformal quantum mechanics on $\M_{K,N}$ admits a dual description as M-theory on asymptotically $\widetilde{AdS}_7\times S^4$ spacetime, with $N$ units of four-form flux on $S^4$, and $K$ units of momentum on the null circle in $\widetilde{AdS}_7$ \cite{Maldacena:2008wh,Dorey:2022cfn}.

This duality---and in particular its implications for black hole microstate counting---was studied in detail in \cite{Dorey:2022cfn}; let us briefly review the results. The generic ultra-spinning black hole in $\widetilde{AdS}_7\times S^4$ should be characterised by a mass $\Delta$, a momentum $K$, two angular momenta $J_\alpha$ transverse to the chosen lightcone in $\widetilde{AdS}_7$, and finally two angular momenta $Q_a$ on $S^4$, which equivalently manifest as electric charges when we reduce to $SO(5)$ gauged supergravity in seven dimensions. This solution should arise in the ultra-spinning limit of the assumed six-parameter Kerr-Newman AdS$_7$ black hole; this solution is unfortunately not known. However, amongst the known solutions is the five-parameter solution of Chow \cite{Chow:2007ts}, which has equal electric charges. By taking the ultra-spinning limit of this solution, in \cite{Dorey:2022cfn} we constructed a five-parameter ultra-spinning $\widetilde{AdS}_7$ black hole, which has equal charges $Q_1=Q_2=Q$. The Bekenstein-Hawking entropy $\mathcal{S}$ was also computed. A full account of this computation is found in \cite{Mouland:2023gcp}.

As in any holographic duality, one must determine the range of parameters for which we can trust the supergravity approximation. It was first found for a simpler AdS$_7$ black hole in \cite{Maldacena:2008wh}, and later reproduced for our black hole in \cite{Dorey:2022cfn}, that the relevant regime is
\begin{align}
  K \gg N^{7/3} \gg 1
  \label{eq: SUGRA regime}
\end{align}
All of $\{\Delta, K, J_\alpha, Q\}$ and the entropy $\mathcal{S}$ scale like $\sqrt{KN^3}$.

The key result of \cite{Dorey:2022cfn} then was the reproduction of the Bekenstein-Hawking entropy of the BPS ultra-spinning black hole from the superconformal index (\ref{eq: LC index QM result}) of the quantum mechanics. The Chow solution admits a BPS limit, in which two degrees of freedom are lost; the mass saturates a BPS bound, while the angular momenta and charges satisfy a non-linear constraint that is familiar for BPS black holes in various dimensions. Following this BPS subspace to the ultra-spinning limit, we thus arrive at a three-parameter BPS ultra-spinning black hole solution in $\widetilde{AdS}_7$, saturating the BPS bound
\begin{align}
  \Delta  = J_1 + J_2 + 4Q
\end{align}
while a particular non-linear relation amongst $\{K,J_1,J_2,Q\}$ is also satisfied\footnote{This non-linear relation is homogeneous in $K$, in the sense that it provides a relationship between $K^{-1/2}J_1, K^{-1/2}J_2$ and $K^{-1/2}Q$.}. 

This BPS black hole corresponds to an ensemble of BPS states in the quantum mechanics with the same charges. Any BPS state carries five charges $\{J_1, J_2, Q_1, Q_2,K\}$. It is useful to choose a new basis $\{L,J_-,Q_-,K,F\}$, where $J_\pm = J_1 \pm J_2$, $Q=Q_1 \pm Q_2$, $L=J_+ + Q_+$ and the fermion number can be taken as $F=-2Q_2$. Then, $\{L, J_-, Q_-,K\}$ span the subspace of charges that commute with the supercharge $\Q$ with respect to which $\I$ is an index; they are respectively conjugate to $\{\epsilon_+, \epsilon_-,m,\beta\}$ in the expression (\ref{eq: 6d LC index def}). Let $d(L,J_-,Q_-,K,F)$ denote the degeneracy of BPS states with these charges, while $\mathcal{C}(L,J_-,Q_-,K)$ is the coefficient of $e^{-\beta K}e^{-\epsilon_+ L} e^{-\epsilon_- J_-}e^{-mQ_-}$ in the Laurent expansion of $\I$. Then,
\begin{align}
  \mathcal{C}(L,J_-,Q_-,K) = \sum_{F\in\mathbb{Z}} (-1)^F d(L,J_-,Q_-,K,F)
\end{align}
The BPS black holes presented in \cite{Dorey:2022cfn} have $Q_-=0$. Furthermore, the non-linear constraint mentioned above is conveniently restated as a determination of $F=F(L,J_-,K)$ as a function of the other charges. Thus, in the regime (\ref{eq: SUGRA regime}) in which we can trust the supergravity solution, we should find at leading order at large $K,N$,
\begin{align}
  \log d\!\left(L,J_-,0,K,F(L,J_-,K)\right) \sim   \mathcal{S}(L,J_-,K) 
\end{align}
where the three charges $L,J_-,K$ uniquely determine the black hole solution, and thus the Bekenstein-Hawking entropy $\mathcal{S}(L,J_-,K)$.

It follows that for some fixed $\{L,J_-,K\}$, the entropy $\mathcal{S}(L,J_-,K)$ must provide a \textit{lower bound} on the growth of the index coefficient $\mathcal{C}(L,J_-,K)$. A priori, one may expect to find more black hole solutions---perhaps with some scalar hair \cite{Markeviciute:2016ivy,Markeviciute:2018yal}---with $Q_-=0$ and $F\neq F(L,J_-,0,K)$, whose entropies also contribute to $\mathcal{C}(L,J_-,0,K)$ and could in principle dominate. However, we found that the growth of the index coefficients saturate this bound; at leading order
\begin{align}
  \log \mathcal{C}\!\left(L,J_-,0,K\right) \sim   \mathcal{S}(L,J_-,K) 
\end{align}
This echoes analogous results in all examples of precision AdS/CFT microstate counting, to the best of our knowledge (e.g. \cite{Benini:2015eyy,Choi:2018hmj,Benini:2018ywd,Cabo-Bizet:2018ehj,Zaffaroni:2019dhb}). It tells us that any new black hole solutions with $F\neq F(L,J_-,K)$ must have Bekenstein-Hawking entropy not more than $\mathcal{S}(L,J_-,K)$, or else that the microstates of such solutions cancel in Bose-Fermi pairs, at least at leading order.

We hope that the holographic models of conformal quantum mechanics described in the present work may provide a useful setting in which to shed light on the role of the ``non-linear'' constraint in the BPS sector of holographic SCFTs.


\section{Four-dimensional $\N=1$ theories}\label{sec: 4d}

In this Section we will apply the general analysis outlined above to the case of $\mathcal{N}=1$ superconformal theories in four dimensions. We can then study the spectrum of local operators by counting states in radial quantisation. Following the notation set out in Section \ref{sec: lens space theory}, states on $S^3$ carry an energy $\hat{H}$ and angular momenta\footnote{We continue to use the convention set out in Part I, whereby $\hat{J}_1,\hat{J}_2$ generate rotations in two orthogonal planes in $\mathbb{R}^4$ in which $S^3$ is embedded, and thus for bosons are both integer, and for fermions both half-integer. The convention used in many of the works we cite in this section (e.g. \cite{Benini:2011nc,Razamat:2013opa}) is to use $j_1,j_2$, related to our generators by $j_1=\frac{1}{2}(\hat{J}_1+\hat{J}_2), j_2 = \frac{1}{2}(\hat{J}_2 - \hat{J}_1)$.} $\hat{J}_1,\hat{J}_2$ in the $SO(4)$ rotation group acting on $S^3$, as well as an R-charge $R$ and charges $Q_a$, $a=1,\dots,r_F$ under the Cartan generators of the flavour group, which we keep generic. The generic partition function then takes the form\footnote{We use $\hat{a}_a$ rather than $\hat{u}_a$ for the global symmetry chemical potentials, to avoid confusion with the fugacities used below}
\begin{align}
  \hat{\Z}(\hat{\mu},\hat{w}_\mu,\hat{v},\hat{v}_a) = \text{Tr}\exp \Big[-\hat{\mu} \hat{H} - \hat{w}_\mu \hat{J}_\mu - \hat{v}R - \hat{a}_a Q_a \Big]
  \label{eq: 4d pf}
\end{align}
These theories have a superconformal index which counts operators lying in short representations of the $\mathcal{N}=1$ superconformal algebra. As usual this corresponds via radial quantisation to an index counting BPS states on $S^{3}$. This, in turn, is a special case of a  supersymmetric $S^{3}$ index which can be defined for any $\mathcal{N}=1$ theory with 
an unbroken $U(1)$ R-symmetry. More precisely, this index counts states of the theory on $S^{3}$ which saturate the BPS bound, 
\begin{eqnarray}
\mathcal{H} & = & \{\mathcal{Q},\mathcal{Q}^{\dagger}\} \,\,=\,\,\hat{H}-\hat{J}_1 - \hat{J}_2 + \frac{3}{2}R\,\,=\,\,0 
\label{eq: 4d BPS bound} 
\end{eqnarray}
The SUSY index is precisely a special case of the $S^3$ partition function (\ref{eq: 4d pf}), when we have
\begin{align}
   -\hat{\mu} + \hat{w}_1 + \hat{w}_2 + 2 \hat{v} = 2\pi i \quad (\text{mod }4\pi i)
  \label{eq: Sigma 4d}
\end{align}
It is straightforward to write the resulting index in the form
\begin{eqnarray}
\hI\!\left(U;p,q \right) & := &  {\rm Tr}\left[(-1)^{F}\exp(-\kappa\mathcal{H})
\,p^{\hat{J}_2-R/2}q^{\hat{J}_1-R/2}\,\prod_{a=1}^{r_{F}} \,u_{a}^{Q_{a}}\right] 
\label{eq: 4d index def}
\end{eqnarray}
Here $U:=\{u_{a}\}$ are fugacities for the flavour symmetries. In the special case where the theory is superconformal, $R$ is the superconformal R-charge and $\hat{H}$ can be identified with the conformal dimension in radial quantisation. In this case the index reduces to the $\mathcal{N}=1$ superconformal index. 

\subsection{The lens space index}

Following \cite{Benini:2011nc,Razamat:2013opa}, we can generalise the index by replacing $S^{3}$ with the lens space, 
\begin{eqnarray}
\lens(\r,1) & := & S^{3} / \mathbb{Z}_{\r}
\nonumber 
\end{eqnarray}
This is achieved by orbifolding the theory by the action of
\begin{align}
   \exp \left(\frac{2\pi i}{\r}(\hat{J}_1 - \hat{J}_2)\right)
\end{align}
This is a \textit{good} orbifold, since all states have integer charge under $(\hat{J}_1 - \hat{J}_2)$, and additionally it preserves the supercharge with respect to which $\I$ is an index. Hence, the index (\ref{eq: 4d index def}) is generalised to an index $\hI_\r$ taking the exact same form, just with the trace taken over states on $\lens(\r,1)$.

In the Hamiltonian framework, $\hI_\r$ can be constructed in the standard way by projecting onto the $\mathbb{Z}_{\r}$ invariant sector of the $S^{3}$ spectrum. One must also introduce a sum over appropriate twisted sectors. The resulting index was computed by localisation in \cite{Razamat:2013opa}. We begin by describing this result. \\

In the following we will consider a generic supersymmetric gauge theory with semi-simple gauge group $G$ of rank $r_{G}$ and flavour symmetry of rank $r_{F}$ as above. The matter content of the theory consists of a $\mathcal{N}=1$ vector multiplet $V$ and $N_{\chi}$ chiral multiplets of R-charges $R_{l}$, $l=1,2,\ldots,N_{\chi}$, which form an (in general  reducible) representation $R_{\chi}$ of $G$ and with charge $\mu_{a}^{(l)}$ under the Cartan generators of the flavour group. The index is formulated as a path 
integral on $\lens(\r,1)\times S^{1}$ with supersymmetry preserving boundary conditions. For $\r>1$, the spatial manifold is no longer simply connected and thus any gauge theory on this space must include sectors of non-trivial holonomy/flux. The resulting index is then  written as a sum over theses sectors. 
\begin{eqnarray}
\hI_{\r}\!\left(U;p,q\right) & = & \sum_{\mathbf{m}\in (\mathbb{Z}_{\r})^{r_{G}}}\,\,
\hI^{(\mathbf{m})}_{\r}\!\left(U;p,q\right)
\label{eq: 4d flux sum}
\end{eqnarray}   
 The resulting contribution to the index is given as follows,  
\begin{eqnarray}
\hI^{(\mathbf{m})}_{\r}\!\left(U;p,q\right) & = & \frac {\hI_{V}(0,1)^{r_{G}}}{\left|\mathcal{W}_{\mathbf{m}}\right|}\, \left(\prod_{i=1}^{r_{G}} \, 
\frac{1}{2\pi i} \,\oint\, \frac{dz_{i}}{z_{i}}\,\right)  \mathcal{M}(Z,\mathbf{m})  \,\,\times \,\, \nonumber \\ && \prod_{\alpha\in \Delta_{+}}\left[ \, \hI_{V}\!\left(\alpha(\mathbf{m}),\exp(\alpha(\mathbf{y}))\right)\,\hI_{V}\!\left(-\alpha(\mathbf{m}),\exp(-\alpha(\mathbf{y}))\right) \right] \times \,\, \nonumber \\ &  &
\prod_{l=1}^{N_{\chi}}\, \hI^{(R_{l})}_{\chi}\!\left(\rho_{l}(\mathbf{m}), 
\exp(\rho_{l}(\mathbf{y}))\prod_{a=1}^{N_{F}} \, u_{a}^{\mu^{(l)}_{a}}\right) 
\label{big} 
\end{eqnarray}
Here we have introduced fugacities $Z:=\{z_{i}\}$, for the Cartan generators of $G$. 
We also set $z_{i}=\exp(y_{i})$ for $i=1,2,\ldots,r_{G}$ and use the vector notation 
$\mathbf{y}:=(y_{1},\ldots, y_{r_{G}})$. In the sector with discrete flux $\mathbf{m}$, the gauge group $G$ is broken to a subgroup $H_{\mathbf{m}}$. In the above expression, $\mathcal{M}(Z,\mathbf{m})$ denotes the Haar measure on the unbroken gauge group and $|\mathcal{W}_{\mathbf{m}}|$ the volume of the corresponding unbroken Weyl subgroup. \\

The second line in (\ref{big}) corresponds to the contribution of the off-diagonal part of the vector multiplet $V$. 
The product is over the set $\Delta_{+}$ of positive roots of $\mathfrak{g}=\mathbb{L}[G]$. The specific contribution of each root is determined by the function, 
\begin{eqnarray}
\hI_{V}\!\left(m,u\right) & := &    
\frac{\hI_{V}^{(0)}\left(m,u\right)}{(1-u^{-1})^{\delta_{m,0}}} \,\times \, 
\left[\Gamma\left(q^{m}u^{-1};q^{\r},pq\right)\Gamma\left(p^{\r-m}u^{-1};p^{\r},pq\right)\right]^{-1} 
\nonumber 
\end{eqnarray}
where $\Gamma(z;p,q)$ is the elliptic Gamma function and  
\begin{eqnarray}
\hI^{(0)}_{V}\!\left(m,u\right) & := &   \left((pq)^{\frac{1}{2}}u^{-1}
\right)^{-\frac{m(\r-m)}{2\r}}\,\times \,\left(\frac{q}{p}\right)^{\frac{m(\r-m)(\r-2m)}{12\r}}
\nonumber 
\end{eqnarray}
Similarly the third line of (\ref{big}) corresponds to the contribution to the index of the chiral multiplets $\chi_{l}$, $l=1,\ldots,N_{\chi}$. Here we diagonalise the action of the Cartan subalgebra $\mathfrak{h}\subset \mathfrak{g}$ on the representation $R_{\chi}$, 
and $\rho_{l}\in \mathfrak{h}^{*}$ is the corresponding weight. The contribution of each weight is determined by the function,  
\begin{eqnarray}
\hI^{(R)}_{\chi}\left(m,u\right) & := &    
\hI_{\chi}^{(R,0)}\left(m,u\right)\,\times \, 
\Gamma\left((pq)^{\frac{R}{2}}q^{\r-m}u;q^{\r},pq\right)\Gamma\left((pq)^{\frac{R}{2}}p^{m}u;p^{\r},pq\right) 
\nonumber 
\end{eqnarray}
where $\Gamma(z;p,q)$ is the elliptic Gamma function and  
\begin{eqnarray}
\hI^{(R, 0)}_{\chi}\left(m,u\right) & := &   \left((pq)^{\frac{1-R}{2}}u^{-1}
\right)^{\frac{m(\r-m)}{2\r}}\,\times \,\left(\frac{p}{q}\right)^{\frac{m(\r-m)(\r-2m)}{12\r}}
\nonumber 
\end{eqnarray} 
\subsection{Holomorphic factorisation}

The lens space $\lens(\r,1)$ can be formed by glueing together two subspaces, each diffeomorphic to $S^{1}\times D_{2}$ where $D_{2}$ denotes the hemisphere, along a common boundary torus. In principle the partition function of any local QFT on this space admits a decomposition into the product of partition functions associated with these subspaces, appropriately summed over boundary conditions. For the supersymmetric index this decomposition is particularly powerful because these constituents, known as {\em holomorphic blocks}, are topological allowing a deformation of the metric on the hemisphere which makes the factorisation explicit.   \\   

Following Pasquetti et al, factorisation is exhibited by first performing a modular transformation on the parameters of the index. In particular we set, 
\begin{eqnarray}
p=\exp\left(2\pi i \frac{\omega_{1}}{\omega_{3}}\right) & \qquad{} & q=\exp\left(2\pi i \frac{\omega_{2}}{\omega_{3}}\right)   
\label{pq}
\end{eqnarray} 
and define, 
\begin{eqnarray}
q_{\tau}=\exp(2\pi i\tau) =\exp\left(2\pi i \frac{Q}{\r\omega_{1}}\right) & \qquad{} &   q_{\sigma}=\exp(2\pi i\sigma) =\exp\left(-2\pi i \frac{\omega_{3}}{\r\omega_{1}}\right)
\nonumber 
\end{eqnarray} 
with $Q=\omega_{1}+\omega_{2}$ and introduce block functions, 
\begin{eqnarray}
\mathcal{B}_{V}(\psi,m;\tau,\sigma) & := & \Gamma\left(x;q_{\tau},q_{\sigma}\right)
\nonumber 
\end{eqnarray} 
and 
\begin{eqnarray}
\mathcal{B}^{D}_{\chi}(\psi,m;\tau,\sigma) &:= & \left[ \Gamma\left(q_{\tau}x^{-1};q_{\tau},q_{\sigma}\right)\right]^{-1} 
\nonumber 
\end{eqnarray}  
with, 
\begin{eqnarray}
x & := & \exp\left(2\pi i\left(\psi+\frac{m}{\r}\right)\right)
\nonumber 
\end{eqnarray}
These functions are closely related to the one-loop contributions of the vector and chiral multiplets to the supersymmetric partition function of the theory on $T^{2}\times D_{2}$. 
In this context, the parameters $\tau$ and $\sigma$ are identified as the equivariant parameter of the $D_{2}$ factor and the complex structure parameters of the $T^{2}$ respectively. More precisely, the block functions given are those appropriate for a vector multiplet with Neumann boundary conditions on $D_{2}$ and chiral multiplets with Dirichlet boundary conditions. \\

The parameter $\tau$ can also be thought of as the complex structure of the boundary torus    of an $S^{1}\times D_{2}$ spatial slice and there is a corresponding action of the modular group. For an element, 
\begin{eqnarray}
g=\left(\begin{array}{cc} a & b\\ c & d \end{array}\right) &\in & SL(2,\mathbb{Z})
\nonumber
\end{eqnarray}  
we have,
\begin{eqnarray}
\tau\rightarrow \tilde{\tau}=\frac{a\tau+b}{c\tau+d}, &  & \sigma \rightarrow \tilde{\sigma}=\frac{\sigma}{c\tau+d},\qquad{}\sigma \rightarrow \tilde{\psi}=\frac{\psi}{c\tau+d},\qquad{}
\nonumber 
\end{eqnarray}
We also define the involution, 
\begin{eqnarray}
\iota:\,\,\{\tau,\sigma,\psi\} & \rightarrow & \{-\tau,-\sigma,-\psi\}
\nonumber{}
\end{eqnarray}
For each element $g$ and each block function we define, 
\begin{eqnarray}
g\circ \mathcal{B}\left(\psi,m ;\tau,\sigma\right) & := & \mathcal{B}\left(\tilde{\psi},m;\tilde{\tau},\tilde{\sigma}\right)
\nonumber 
\end{eqnarray}
and
\begin{eqnarray}
 \iota\mathcal{B}\left(\psi,m ;\tau,\sigma \right)& := &
\mathcal{B}\left(-\psi,m ;-\tau,-\sigma\right)
\end{eqnarray}
Finally,  lens space $\lens(\r,1)$ can be constructed by glueing together two copies 
of $D^{2}\times S^{1}$ with an appropriate twist of the boundary torus corresponding to the action of the modular transformation $g_{\r}:=ST^{\r}S$, where $S:\tau\rightarrow -1/\tau$ and $T:\tau\rightarrow \tau+1$ denote the standard generators of $SL(2,\mathbb{Z})$. Following \cite{Nieri:2015yia}, we define a corresponding pairing of the block functions, 
\begin{eqnarray}
\left\| \mathcal{B}(\psi,m;\tau,\sigma)  \right\|_{\r}^{2} & = &   
\mathcal{B}\left(\psi,m ;\tau,\sigma \right)\,\cdot\,g_{\r}\circ\iota \mathcal{B}\left(\psi,\r-m ;\tau,\sigma \right)
\label{pair} 
\end{eqnarray}
With these definitions, we can write the lens space index of any consistent theory without gauge anomalies as, 
\begin{eqnarray}
\hI_{\r}\!\left(U;p,q\right) & =& \exp\left(i\pi {\mathcal P}_{\rm gl}+i\pi \mathcal{P}_{\rm gl}^{\rm 3d}\right)\,\,\sum_{\mathbf{m}\in (\mathbb{Z}_{\r})^{r_{G}}}\,\,
\frac{1}{\left|\mathcal{W}_{\mathbf{m}}\right|}\, \left(\prod_{i=1}^{r_{G}} \, 
\frac{1}{2\pi i} \,\oint_{C}\, \frac{dz_{i}}{z_{i}}\,\right)\,\,\exp\left( i\pi \mathcal{P}_{\rm loc}^{\rm 3d}(\mathbf{y})\right)\,\,\left\|\Upsilon^{4d}\right\|_{\r}^{2} \nonumber 
\label{weak}
\end{eqnarray}
where the integrand is written in the factorised form with, 
\begin{eqnarray} 
\Upsilon^{4d}  =  \prod_{\alpha\in\Delta_{+}} \left[\mathcal{B}_{V}(\psi_{\alpha},\alpha(\mathbf{m});\tau,\sigma)\mathcal{B}_{V}(\psi_{-\alpha},-\alpha(\mathbf{m});\tau,\sigma)\right] \,\times \prod_{l=1}^{N_{\chi}} 
\mathcal{B}^{D}_{\chi}(\psi_{l},\rho_{l}(\mathbf{m});\tau,\sigma)
\label{ups} 
\end{eqnarray}
where the arguments of the block functions are given as, 
\begin{eqnarray}
\psi_{\alpha}=-\frac{\sigma}{2\pi i}\alpha(\mathbf{y}), & {\rm and} & \psi_{l}=-\frac{\sigma}{2\pi i}\left(\rho_{l}(\mathbf{y})+\sum_{a=1}^{r_{F}}\mu_{a}\log u_{a}\right) +\frac{\tau}{2}\, R_{l} 
\label{args}
\end{eqnarray}
for $l=1,\ldots,r_{F}$ and the pairing is applied factorwise to the whole integrand with, 
\begin{eqnarray}
\left\|\Upsilon^{4d}\right\|_{\r}^{2} & := & \prod_{\alpha\in\Delta_{+}} \left\|\mathcal{B}_{V}(\psi_{\alpha},\alpha(\mathbf{m});\tau,\sigma)\right\|_{\r}^{2}\,\, \left\|\mathcal{B}_{V}(\psi_{-\alpha},-\alpha(\mathbf{m});\tau,\sigma)\right\|_{\r}^{2} \nonumber \\  \,&  \times & \qquad{} \prod_{l=1}^{N_{\chi}} 
\left\|\mathcal{B}^{D}_{\chi}(\psi_{l},\rho_{l}(\mathbf{m});\tau,\sigma)\right\|_{\r}^{2}
\nonumber 
\end{eqnarray}
We note that the quantity $\Upsilon^{4d}$ depends on the gauge fugacities $\mathbf{y}$ and the fluxes $\mathbf{m}$ only via the holomorphic combinations, 
\begin{eqnarray}
s_{i} & := & \exp\left(\frac{2\pi i}{r\omega_{1}}y_{i}+\frac{2\pi i}{r}m_{i}\right) 
\nonumber 
\end{eqnarray}
for $i=1,\ldots,r_{G}$, which we will denote collectively as $S$. In the following it will be useful to consider the factorised integrand as a function of these variables: 
$\Upsilon^{4d}=\Upsilon^{4d}(S;q_{\tau},q_{\sigma})$.\\

The prefactors in (\ref{weak}) are exponentials of certain polynomials $\mathcal{P}_{\rm gl}$ and $\mathcal{P}_{\rm 3d}$ in the fugacities $U$ which are determined by the global anomalies of the four dimensional theory and the parity anomaly of its corresponding three dimensional reduction. The integrand also includes the exponential of a polynomial $\mathcal{P}^{\rm 3d}_{\rm loc}$ which is quadratic in the gauge and global fugacities encoding the induced Chern-Simons terms for the gauge field. This term can be also by factorised in terms of $\theta$-functions with modular parameter $\tau$ which can then be absorbed into the definition of the blocks. There is also an additional ambiguity in defining the blocks which originates in the existence of ratios of $\theta$ functions which are elliptic on the boundary torus and glue to unity under the pairing (\ref{pair}). 
For more details see \cite{Nieri:2015yia}. \\

Finally we can define a basis of holomorphic blocks as contour integrals of the factorised integrand. The blocks span a vector space whose dimension is equal to the number $N_{\rm Vac}$ of massive vacua of the theory. Different basis elements, $\mathbb{B}_{\alpha}$,   labelled by an index $\alpha=1,\ldots,N_{\rm Vac}$ are selected by corresponding integration contours $\Gamma_{\alpha}$ on the complexified maximal torus of $G$. Schematically,  
\begin{eqnarray}
\mathbb{B}_{\alpha}[U;\tau,\sigma] & = & \oint_{\Gamma_{\alpha}}\,\,\prod_{i=1}^{r_{G}}\, 
\frac{ds_{i}}{2 \pi i s_{i}}\,\,\Upsilon^{4d}
\end{eqnarray}
Here we assume that any induced Chern-Simons terms for the gauge field have been factorised in $\theta$ functions and reabsorbed in the definition of the $\Upsilon^{4d}$. The remaining ambiguity in defining the blocks can also be understood as a change of basis in space of blocks. \\

In favourable cases, a strong version of factorisation holds and the index can be expressed  as, 
\begin{eqnarray}
\hI_{\r}\left(U;p,q\right) & =& \exp\left(i\pi {\mathcal P}_{\rm gl}+i\pi \mathcal{P}_{\rm gl}^{3d}\right)\,\,\sum_{\mathbf{m}\in (\mathbb{Z}_{\r})^{r_{G}}}\,\,
\frac{1}{\left|\mathcal{W}_{\mathbf{m}}\right|}\, \left(\prod_{i=1}^{r_{G}} \, 
\frac{1}{2\pi i} \,\oint_{C}\, \frac{dz_{i}}{z_{i}}\,\right)\,\,\left\|\Upsilon^{4d}\right\|_{\r}^{2} \nonumber \\ 
& = & \exp\left(i\pi {\mathcal P}_{\rm gl}+i\pi \mathcal{P}_{\rm gl}^{3d}\right)\,\,\sum_{\alpha=1}^{N_{\rm Vac}}\,\,  \left\| \mathbb{B}_{\alpha}\right\|_{\r}^{2}
\nonumber
\end{eqnarray}
where the pairing is the same one defined in (\ref{pair}) above. Thinking of the blocks as an $N_{\rm Vac}$-component vector $\vec{\mathbb{B}}$, both modular transformations and the involution are represented as linear transformations. In particular we define a representation $\mathcal{R}$ of $SL(2,\mathbb{Z})$ of dimension $N_{\rm Vac}$ with generators, 
\begin{eqnarray}
\mathbb{S}:=\mathcal{R}(S), & & \mathbb{T}:=\mathcal{R}(T)
\end{eqnarray}
the involution $\iota$ also lifts to a linear transformation acting on the blocks. 
Finally the index takes the elegant form, 
\begin{eqnarray}
\hI_{\r} & = & \vec{\mathbb{B}}^{T}\cdot \mathbb{S}\mathbb{T}^{\r}\mathbb{S}\cdot 
\iota \vec{\mathbb{B}}
\nonumber
\end{eqnarray}
We emphasise that our analysis does not require this stronger form of holomorphic factorisation, but only the factorisation (\ref{weak}) of the integrand.  
\subsection{Deriving the DLCQ index}

Following Section \ref{sec: lightcone theory}, we now consider the same $\N=1$ theory now formulated on the null-compactified pp-wave (\ref{eq: pp-wave}). The general partition function of the resulting theory takes the form
\begin{align}
  \Z(\beta,\mu,w_\alpha,u_a) = \text{Tr}_{\text{LC}} \exp \Big[ - \beta K - \mu \Delta - w J - v R  - a_a Q_a \Big]
  \label{eq: 4d lc pf}
\end{align}
with discrete null momentum $K$, Hamiltonian $\Delta$ generating $x^-$ translations, and single rotation $J$ in the plane transverse to the lightcone. The supercharge $\Q$ appearing in the BPS bound (\ref{eq: 4d BPS bound}) is preserved by the null compactification, and gives rise to the BPS bound
\begin{align}
  \{\Q,\Q^\dagger\} = \H =  \Delta - J + \frac{3}{2}R
\end{align}
The DLCQ partition function $\Z$ is then promoted to an index for this BPS bound on the chemical potential subspace
\begin{align}
  -2\mu + w + 2v = 2\pi i\quad (\text{mod } 4\pi i) 
\end{align}
The resulting index is easily shown to take the general form
\begin{align}
  \I(U;\tilde{p},q) = \text{Tr}_{\text{LC}}\left[(-1)^{F}\exp(-\kappa\mathcal{H})
\,\tilde{p}^{K}q^{J-R/2}\,\prod_{a=1}^{r_{F}} \,u_{a}^{Q_{a}}\right] 
\label{eq: 4d DLCQ index def}
\end{align}
Then, the general proposal (\ref{eq: general pf limit}) implies that we can compute the DLCQ index $\I$ from the lens space index $\hI_\r$ as\footnote{This is straightforwardly deduced by comparing (\ref{eq: 4d index def}) with (\ref{eq: 4d DLCQ index def}) using the relations (\ref{eq: field theory charge rescaling}). Necessarily, one can show that this limit is a supersymmetry-preserving limit of the form (\ref{eq: limit of chemical potentials}). }
\begin{align}
  \I(U;\tilde{p},q) = \lim_{\r\to\infty} \hI_\r\!\left(U;p = \tilde{p}^{1/\r}, q\right)
  \label{eq: 4d index limit}
\end{align}
So let us consider the behaviour of $\hI_\r(U;p, q)$ as we approach $\r\to\infty$ with $p^\r = \tilde{p}$ and $q$ held fixed. The limit is complicated by the behaviour of the flux sum in (\ref{eq: 4d flux sum}) whose range diverges as $\r\rightarrow 0$. An important issue is whether the contributions of the configurations of large flux $m\sim O(\r)$ dominate in the limit. To investigate this question we scale the summation variables by defining $\mathbf{m}=\r\mathbf{x}$, where each component of the continuous real variable $\mathbf{x}$, is subject to the identification $x_{i}\sim x_{i}+1 $. In this limit, the sum over $\mathbf{m}$ in (\ref{big}) becomes an integral over $\mathbf{x}$ and the leading behaviour of the integrand is determined by the functions $\mathcal{I}^{(0)}_{V}$ and $\mathcal{I}^{(R,0)}_{\chi}$. In particular we find that implementing the limit (\ref{eq: 4d index limit}), the leading behaviour of the index is,   
\begin{eqnarray}
\hI_{\r} & \sim &  \r^{r_{G}}\,\,\hI_{\infty}\,\,\int d^{r_{G}}\mathbf{x}\,\,\, \exp\left(\frac{\r}{2}\log q\,\cdot\,L_{h}(\mathbf{x})\right)
\label{asymp}
\end{eqnarray}
where $\hI_{\infty}=\hI_{\infty}(U,\tilde{p}, q)$ is of order one. 
The function $L_{h}$ is given explicitly as, 
\begin{eqnarray}
L_{h}(\mathbf{x}) & = & \sum_{l=1}^{N_{\chi}}\, \frac{1}{2}\left(1-R^{\rm eff}_{l}\right) 
\vartheta\left(\rho_{l}(\mathbf{x})\right)\,\,-\,\,
\sum_{\alpha\in \Delta_{+}}\vartheta\left(\alpha(\mathbf{x})\right) 
\nonumber{}
\end{eqnarray}
where we define the action, 
\begin{eqnarray}
\vartheta(x) & := & \{x\}\left(1-\{x\}\right) 
\nonumber 
\end{eqnarray}
with $\{x\}=x-\lfloor x \rfloor$ and 
\begin{eqnarray} 
R_{l}^{\rm eff} & := & R_{l}-\frac{2}{\log q}\sum_{a=1}^{r_{F}}\,\mu_{a}^{(l)}\log u_{a} 
\nonumber 
\end{eqnarray}
As the exponent in the integrand of (\ref{asymp}) grows linearly with $\r$, the remaining integral over $\mathbf{x}$ can be evaluated by saddle point and the leading behaviour of 
the index in the integrand is therefore governed by the stationary points of the function $L_{h}(\mathbf{x})$. Remarkably, for any given theory, exactly the same integral arises in the analysis of the high-temperature limit of the $S^{3}$ index \cite{ArabiArdehali:2015ybk}. In this context, $L_{h}$ is known as the {\it Rains function} of the theory and its stationary points similarly govern the high-temperature asymptotics of the index. This striking relation between the limiting behaviours of the index on two different spaces has a natural explanation as a consequence of the modular symmetry proposed by Shaghoulian in \cite{Shaghoulian:2016gol}. The modular properties in question belong to the SUSY partition function; 
\begin{eqnarray}
Z_{\rm SUSY} \!\left(S^{3}/\mathbb{Z}_{\r}\times S^{1}\right) & := & 
\exp\left(-\beta\mathcal{E}_{\rm SUSY}\right)\,\,\hI_{\r}\!\left(U;p,q\right)
\nonumber{} 
\end{eqnarray}
which is related to the lens index $\hI_{\r}$ by an exponential factor involving the {\em supersymmetric Casimir energy}\footnote{See for instance \cite{Assel:2015nca} for a useful discussion of this object.} $\mathcal{E}_{\rm SUSY}$. Here $p$ and $q$ are parameterised as in (\ref{pq}) and we identify the ``temperature'' $\beta:=2\pi/\omega_{3}$. In \cite{Shaghoulian:2016gol}, evidence is presented in favour of the duality,   
 \begin{eqnarray}
Z_{\rm SUSY} \left(S^{3}/\mathbb{Z}_{\r}\times S^{1}\right) & = &   Z_{\rm SUSY} \left(S^{3}\times S^{1}/\mathbb{Z}_{\r}\right) 
\nonumber
\end{eqnarray}
for $\r>>1$. Comparing Ardehali's formula (3.9) \cite{ArabiArdehali:2015ybk} and the asymptotic behaviour  
(\ref{asymp}), we see that our results are indeed consistent with the proposed duality\footnote{See also Appendix B of \cite{ArabiArdehali:2019zac} for a related discussion.}. 

According to the above discussion, the limiting behaviour of the index depends strongly on the properties of the Rains function $L_{h}$. 
Fortunately, for a large class of theories discussed in \cite{ArabiArdehali:2015ybk}, the resulting behaviour is very simple. For this class of theories, the following property holds, 
\begin{enumerate}[label=\textbf{\Roman*:}]
  \item The Rains function $L_{h}(\mathbf{x})$ has an isolated global minimum at the origin $\mathbf{x}=0$. We also have $L_{h}(0)=0$
\end{enumerate}
 For theories outside this class, where the minimum occurs at a non-zero value of $\mathbf{x}$, one cannot obtain the DLCQ index by applying our limit in a straightforward way. Plausibly this behaviour corresponds to a condensation of the discrete holonomy modes which takes us away from the conformal phase of the theory. 
As explained in \cite{ArabiArdehali:2015ybk}, the stated property is closely aligned with the validity of the di Pietro-Komargodski formula for the high-temperature asymptotics of the SUSY partition function \cite{DiPietro:2014bca}. Interestingly, the special case of $\mathcal{N}=4$ supersymmetric Yang-Mills theory is marginal as the corresponding Rains function vanishes identically. In that case there is a non-trivial moduli space of minima which should be integrated over. We will not analyse this in detail here.\\

For theories obeying property $\mathbf{I}$ we may complete the analysis and find a finite limiting value for the lens space index, $\hI_{\r}\rightarrow \I$ in the limit (\ref{eq: 4d index limit}). Ignoring the constant exponential prefactors associated with global anomalies we have,  
\begin{eqnarray}
\I(U;\tilde{p},q) = \hI_{\infty}(U;\tilde{p},q)  =  \,\,\prod_{i=1}^{r_{G}}\,\,\oint_{C}\,\, 
\frac{dz_{i}}{2 \pi i z_{i}}\,\,\, \Xi^{4d}
\nonumber
\end{eqnarray}
with 
\begin{eqnarray}
\Xi^{4d} & = & \prod_{\alpha\in\Delta_{+}} \left[\Gamma\left(\tilde{p}\exp(-\alpha(\mathbf{y}));\tilde{p},q\right)\Gamma\left(\tilde{p}\exp(+\alpha(\mathbf{y}));\tilde{p},q\right)\right]^{-1}
 \,\times \prod_{l=1}^{N_{\chi}} \Gamma\left(q^{R_{l}^{\rm eff}/2}\exp(-\rho_{l}(\mathbf{y}));\tilde{p},q\right)
\nonumber 
\end{eqnarray}
Up to $\theta$ functions corresponding to induced Chern-Simons terms for the gauge field at integer level, we  can then write the DLCQ index as a holomorphic block of the form, 
\begin{eqnarray}
\I(U;\tilde{p},q) & = & \oint_{\Gamma_{\infty}}\,\,\prod_{i=1}^{r_{G}}\, 
\frac{d{z}_{i}}{2 \pi i {z}_{i}}\,\,\hat{\Upsilon}^{4d}
\end{eqnarray}
with $\hat{\Upsilon}^{4d}={\Upsilon}^{4d}(Z;\tilde{p},q)$. The contour $\Gamma_{\infty}$ corresponds to the unit circle in each of the variables $z_{i}$. 
With the specified choice of contour, 
the resulting expression precisely corresponds to a particular choice of holomorphic block.\\

In conclusion, for the class of $\mathcal{N}=1$ theories obeying condition $\mathbf{I}$   
our limit provides a prediction for the DLCQ index. It coincides with a particular holomorphic block corresponding to a supersymmetric partition function for the same  theory on $T^{2}\times D_{2}$ with complex structure parameters $\hat{\tau}:=\log\tilde{p}/2\pi i$ and $\hat{\sigma}:=\log q/2\pi i$. As above the relevant boundary conditions on the hemisphere are Neumann for the vector multiplet and Dirichlet for the chiral multiplets.  

\newpage
{\noindent \LARGE\bf\color{mygreen}  Appendices}	
\addcontentsline{toc}{section}{\color{mygreen}\large Appendices}
\appendix

\section{Schr\"odinger algebra}\label{app: algebra}

The conformal isometries of $d$-dimensional Minkowski space
\begin{align}
  \tilde{ds}^2 = -2 d\tilde{x}^+ d\tilde{x}^- + d\tilde{x}^i d\tilde{x}^i
  \label{eq: Minkowski again}
\end{align}
with $i=1,\dots,d-2$, form $SO(2,d)$. The stabiliser of the null translation $\tilde{x}^+$ within $\frak{so}(2,d)$ defines the Schr\"odinger group $\frak{Schr}(d-2)\subset \frak{so}(2,d)$ in $(d-2)$ spatial dimensions. Let us set our conventions for the generators of $\frak{Schr}(d-2)$. Descending from the Poincar\'{e} subalgebra of $\frak{so}(2,d)$, we have a Hamiltonian $H$, spatial translations $P_i$, spatial rotations $M_{ij}$, a Galilean boost $G_i$, and finally a particle number $K$. Descending from the additional conformal generators in $\frak{so}(2,d)$, we additionally find a Lifshitz dilatation $D$, and a `special' conformal generator $C$. These generators $G$ correspond to the following conformal Killing vectors $G_\partial$ of $\tilde{ds}^2$,
\begin{align}
  	i(H)_\partial 		&= \tilde{\partial}_-	& i(K)_\partial 		&= \tilde{\partial}_+	\nn\\[-0.3em]
  	i(P_i)_\partial 	&= \tilde{\partial}_i	& i(D)_\partial 		&= 2\tilde{x}^- \tilde{\partial}_- + \tilde{x}^i \tilde{\partial}_i	\nn\\[-0.3em]
  	i(M_{ij})_\partial 	&= \tilde{x}^i\tilde{\partial}_j	 - \tilde{x}^j\tilde{\partial}_i	& 	i(C)_\partial 		&= \tfrac{1}{2}\tilde{x}^i\tilde{x}^i \tilde{\partial}_+ + (\tilde{x}^-)^2 \tilde{\partial}_- + \tilde{x}^- \tilde{x}^i \tilde{\partial}_i	\nn\\[-0.3em]
  	i(G_i)_\partial 	&= \tilde{x}^i \tilde{\partial}_+ + \tilde{x}^- \tilde{\partial}_i	
\end{align}
from which one can deduce the commutation relations of $\frak{Schr}(d-2)$.

\section{Instanton partition functions}\label{app: 6d conventions}

Here, we give details of the instanton partition function
\begin{align}
  \Z_\text{inst}(\beta,\epsilon_1,\epsilon_2,m,\lambda_I) = \sum_{K=0}^\infty e^{-\beta K} \Z_\text{inst}^{(K)}(\epsilon_1,\epsilon_2,m,\lambda_I)
\end{align}
which appears in both the lightcone index (\ref{eq: LC index QM result}) and lens space index (\ref{eq: IL explicit result}) of the six-dimensional $U(N)$ $(2,0)$ theory. As discussed in Section \ref{subsec: 6d result from QM}, $\Z_\text{inst}$ is precisely the non-perturbative contribution to the Nekrasov partition function of an $\N=1^*$ $SU(N)$ super-Yang-Mills theory in five dimensions - for further details of this identification, see \cite{Dorey:2022cfn}.\\

We first set up some notation. A partition $\ptt$ is a finite sequence of non-increasing positive integers $\ptt_1\geq \ptt_2\geq \ptt_3... \geq \ptt_{\ell(\ptt)}>\ptt_{\ell(\ptt)+1}=0$. Here $\ell(\ptt)$ is called the length of the partition and $|\ptt|=\sum_{p}\ptt_{p}$ is its weight. The set of partitions is denoted $\mathcal{P}$. 

One usually visualises a partition $\ptt$ by drawing the associated Young diagram $Y(\ptt)$. A box $s$ in a young diagram $Y(\ptt)$ is labelled by its coordinates $(p,q)$ where $p = 1,..., l(\ptt)$. Arm and leg lengths of a box $s=(p,q)\in Y(\ptt)$  
are defined as,  

\begin{eqnarray}\label{leg_arm_length}
    a(s) = \ptt_p - q, & \qquad{} & l(s)= \ptt^{\vee}_q - p
\end{eqnarray}
where the dual $\ptt^{\vee}$ of a partition $\ptt$ is obtained by interchanging rows and columns in the Young diagram. This definition can extended to boxes $s=(p,q)$ which lie outside the Young diagram.\\

We also consider $N$-component vectors of partitions denoted $\vec{\ptt}=(\ptt^{(1)},\ptt^{(2)},\ldots,\ptt^{(N)})$ with total weight, 
\begin{eqnarray}
||\vec{\ptt}|| & = & \sum_{i=1}^{N}\, |\ptt^{(i)}|
\label{tot}
\end{eqnarray}
Such vectors of partitions, along with their associated vectors of Young tableaux, are referred to as $N$-coloured. The set of all $N$-coloured partitions is denoted $\mathcal{P}^N$.

For each box $s\in Y(\ptt^{(I)})$ we define, 
\begin{eqnarray}
g_{IJ}(s) & = & -a_{I}(s)+l_{J}(s) \nonumber \\ 
f_{IJ}(s) & = & -a_{I}(s)-l_{J}(s)-1
\label{fg}
\end{eqnarray}
Here $a_{I}(s)$ and $l_{J}(s)$ are the arm and leg lengths of box $s$, relative to the Young Tableaux $Y(\ptt^{(I)})$ and $Y(\ptt^{(J)})$ respectively. 

We define a  Plethystic exponential for any function $f$ of $r$ formal variables $\{x_{1},x_{2},\ldots,x_{r}\}$ as, 
\begin{eqnarray}
{\rm Pexp}\left[f(x_{1},x_{2},\ldots,x_{r})\right) & = & 
\exp \left[\sum_{n\geq 1} \frac{1}{n} f\left(x^{n}_{1},x^{n}_{2},\ldots,x^{n}_{r}\right) \right]
\nonumber{} 
\end{eqnarray}
For any $X$ we also define, 
\begin{eqnarray}
\left[ X\right] & := & \sqrt{X}-\frac{1}{\sqrt{X}} 
\nonumber 
\end{eqnarray}
Then, for each $K$, $\Z_\text{inst}^{(K)}$ is given by a sum over all $N$-coloured Young tableaux of total weight $K$, as 
\begin{eqnarray}
\mathcal{Z}^{(K)}_{\rm inst}(\epsilon_1,\epsilon_2,m,\lambda_I) & = & \sum_
{\vec{\ptt}\in \mathcal{P}^{N}: ||\vec{\ptt}||=K} \,\, \prod_{I,J=1}^{N}\,\, \prod_{s\in Y(\ptt^{(I)})}\,\,{\rm Pexp}\left(\frac{z_{I}}{z_{J}}\,t^{g_{IJ}(s)}\,x^{f_{IJ}(s)}\left[ty\right] \left[t/y\right]\right) \nonumber \\
\label{inst2}
\end{eqnarray}
where
\begin{align}
  t = e^{-\epsilon_+},\qquad x = e^{-\epsilon_-},\qquad y = e^{-m},\qquad z_I = e^{-i\lambda_I}
\end{align}

\section{Details of 6d $(2,0)$ index limit}\label{app: 6d limit}

In this Appendix, we provide full details of the computation that derives the lightcone index (\ref{eq: 6d LC index def}) of the 6d $(2,0)$ theory from its lens space index (\ref{eq: 6d lens index}), in the limit
\begin{align}
  \lim_{L\to\infty} \hI_L\!\left(\hat{\beta}=\tfrac{2}{3}\epsilon_+ + \tfrac{\beta}{L},\,\hat{v}_\alpha=\epsilon_\alpha - \tfrac{2}{3}\epsilon_+,\,\hat{m}=m\right)
\end{align}
%
Let us first decompose $\hI_L$ into the contributions from each flux sector, as
\begin{align}
  \hI_L(\hat{\beta},\hat{v}_\alpha,\hat{m}) = \sum_{s_1,\dots,s_N=-\infty}^\infty \hI_L^{(s)}(\hat{\beta},\hat{v}_\alpha,\hat{m})
\end{align}
Then, assuming
\begin{align}
  \text{Re}(\epsilon_{1,2})>0\ ,
  \label{eq: eps constraint}
\end{align}
we will show the following two results:  

\begin{enumerate}
  \item In Section \ref{subsec: non-zero flux}, we show that
  \begin{align}
  \lim_{L\to\infty} \hI^{(s)}_L\!\left(\hat{\beta}=\tfrac{2}{3}\epsilon_+ + \tfrac{\beta}{L},\,\hat{v}_\alpha=\epsilon_\alpha - \tfrac{2}{3}\epsilon_+,\,\hat{m}=m\right) = 0
  \label{eq: 6d limit flux result}
\end{align}
for all $s\neq (0,\dots,0)$.
\item In Section \ref{subsec: zero flux}, we show that
\begin{align}
  \lim_{L\to\infty} \hI^{(s=(0,\dots,0))}_L\!\left(\hat{\beta}=\tfrac{2}{3}\epsilon_+ + \tfrac{\beta}{L},\,\hat{v}_\alpha=\epsilon_\alpha - \tfrac{2}{3}\epsilon_+,\,\hat{m}=m\right) = \I(\beta,\epsilon_\alpha,m)
    \label{eq: 6d limit zero flux result}
\end{align}
for $\I(\beta,\epsilon_\alpha,m)$ as in (\ref{eq: LC index QM result}).
\end{enumerate}
Combined, these two results gives us the desired result, that
\begin{align}
  \lim_{L\to\infty} \hI_L\!\left(\hat{\beta}=\tfrac{2}{3}\epsilon_+ + \tfrac{\beta}{L},\,\hat{v}_\alpha=\epsilon_\alpha - \tfrac{2}{3}\epsilon_+,\,\hat{m}=m\right)= \I(\beta,\epsilon_\alpha,m)
\end{align}
where $\I(\beta,\epsilon_\alpha,m)$ is as given in (\ref{eq: LC index QM result}).

\subsection{Instanton decomposition}

It will be useful to further decompose the fixed-flux index $\hI_L^{(s)}$ by expanding each of the three copies of $\Z_\text{inst}$ appearing in $\Z^{(s)}_\text{non-pert}$ into contributions of fixed instanton number, as in (\ref{eq: total non-pert Nek}). We find
\begin{align}
  &\hI_L^{(s)}\!\left(\hat{\beta}=\tfrac{2}{3}\epsilon_+ + \tfrac{\beta}{L},\,\hat{v}_\alpha=\epsilon_\alpha - \tfrac{2}{3}\epsilon_+,\,\hat{m}=m\right) \nn\\
  &\qquad = \sum_{K_1,K_2,K_3=0}^\infty e^{-\beta(K_1+K_2+K_3)} e^{-LK_1 \epsilon_1}e^{-LK_2 \epsilon_2} \,\hI_{L;K_1,K_2,K_3}^{(s)}(\beta,\epsilon_\alpha,m)
\end{align}
where
\begin{align}
  \hI_{L;K_1,K_2,K_3}^{(s)}(\beta,\epsilon_\alpha,m)  &= \frac{1}{N!}\oint \prod_{I=1}^N \left(\frac{d\lambda_I}{2\pi}\right) e^{\frac{1}{2}\beta \sum_I s_I^2 }\exp\left(-L\sum_I \left(-\frac{1}{3}\epsilon_+ s_I^2 + i s_I \lambda_I\right)\right)\nn\\
   &\hspace{40mm}\times \Z^{(s)}_\text{pert}\! \left(\hat{\beta}=\tfrac{2}{3}\epsilon_+ + \tfrac{\beta}{L},\,\hat{v}_\alpha=\epsilon_\alpha - \tfrac{2}{3}\epsilon_+,\,\hat{m}=m ; \lambda\right)\nn\\
   &\hspace{40mm}\times \prod_{\mu=1}^3 \Z_\text{inst}^{K_\mu} \left(\epsilon_1^{(\mu)},\epsilon_2^{(\mu)},m^{(\mu)},\lambda_I^{(\mu)}\right) 
   \label{eq: ILKK def}
\end{align}
Using (\ref{eq: Nek parameters of each factor}), we have
\begin{align}
  \epsilon_1^{(1)} 	&= \epsilon_2-\epsilon_1										&  
  \epsilon_1^{(2)} 	&= -\epsilon_2												&
  \epsilon_1^{(3)} 	&= \epsilon_1													\nn\\	
  \epsilon_2^{(1)} 	&= -\epsilon_1												&  
  \epsilon_2^{(2)} 	&= \epsilon_1-\epsilon_2										&
  \epsilon_2^{(3)} 	&= \epsilon_2													\nn\\			
  m^{(1)}			&= m+\tfrac{1}{2}\epsilon_1 + \tfrac{\beta}{2L}				& 
  m^{(2)} 			&= m+\tfrac{1}{2}\epsilon_2 + \tfrac{\beta}{2L}				&
  m^{(3)} 			&= m + \tfrac{\beta}{2L}					\nn\\	
  \lambda_I^{(1)} 	&= \lambda_I + \tfrac{i}{3}(2\epsilon_1-\epsilon_2)s_I		&  
  \lambda_I^{(2)} 	&= \lambda_I + \tfrac{i}{3}(2\epsilon_2-\epsilon_1)s_I		&
  \lambda_I^{(2)} 	&= \lambda_I - \tfrac{2i}{3}\epsilon_+s_I
\end{align}

\subsection{Vanishing of non-zero flux terms}\label{subsec: non-zero flux}

Let us first show that
\begin{align}
  \lim_{L\to\infty} \hI^{(s)}_{L;K_1,K_2,K_3}(\beta,\epsilon_\alpha,m) = 0
\end{align}
for all $s\neq (0,\dots,0)$ and all $K_\mu$. This, together with (\ref{eq: eps constraint}), implies the desired result (\ref{eq: 6d limit flux result}).\\

So we consider $\hI^{(s)}_{L;K_1,K_2,K_3}(\beta,\epsilon_\alpha,m)$ for some fixed, generic $K_1,K_2,K_3$. Consider then the integral over some such $\lambda_{I}$, keeping fixed the remaining $\lambda_J$. The integrand then has a finite number of poles, around which the contour must navigate. The poles in the perturbative piece are all of one of the forms \cite{Kim:2013nva}
\begin{align}
  \lambda_{I} = \left\{\begin{aligned}
  &\lambda_J + \frac{in}{3} (2\epsilon_1 - \epsilon_2) \pm i\left(m + \frac{1}{3}\epsilon_+ + \frac{\beta}{2L}\right)\quad &(\text{mod }  2\pi)		\\
  &\lambda_J + \frac{in}{3} (2\epsilon_2 - \epsilon_1) \pm i\left(m + \frac{1}{3}\epsilon_+ + \frac{\beta}{2L}\right)\quad &(\text{mod }  2\pi)		\\
  &\lambda_J + \frac{in}{3} (\epsilon_1 + \epsilon_2) \pm i\left(m + \frac{1}{3}\epsilon_+ + \frac{\beta}{2L}\right)\quad &(\text{mod }  2\pi)
\end{aligned}\right.	
  \label{eq: pert poles}
\end{align}
for some $J$ such that $s_{IJ}:= s_{I}-s_J\neq 0$, where $n\in\mathbb{Z}$ is some integer that may take a different value for each pole. Meanwhile, all poles in the non-perturbative piece are of one of the forms
\begin{align}
  \lambda_{I} = \left\{\begin{aligned}
  &\lambda_J - \tfrac{i}{3} s_{IJ}(2\epsilon_1 - \epsilon_2) - i\, \Big( -2\epsilon_- \!\left(a_I(\alpha_1)+\tfrac{1}{2}\right) + \epsilon_1\left(l_J(\alpha_1)+\tfrac{1}{2}\right) \pm \tfrac{1}{2}(2\epsilon_1 - \epsilon_2) \Big)&(\text{mod }  2\pi)		\\
 &\lambda_J - \tfrac{i}{3} s_{IJ}(2\epsilon_2 - \epsilon_1) - i\, \Big( +2\epsilon_- \!\left(a_I(\alpha_2)+\tfrac{1}{2}\right) + \epsilon_2\left(l_J(\alpha_2)+\tfrac{1}{2}\right) \pm \tfrac{1}{2}(2\epsilon_2 - \epsilon_1) \Big)&(\text{mod }  2\pi)		\\
    &\lambda_J + \tfrac{2i}{3} s_{IJ}\epsilon_+ - i\, \Big(\epsilon_1\! \left(a_I(\alpha_3)+\tfrac{1}{2}\right) - \epsilon_2\!\left(l_J(\alpha_3)+\tfrac{1}{2}\right) \pm \epsilon_+ \Big) &(\text{mod }  2\pi)
  \end{aligned}\right.\label{eq: non-pert poles}
\end{align}
These arise from $\Z_{\text{inst}}^{K_1},\Z_{\text{inst}}^{K_2}$ and $\Z_{\text{inst}}^{K_3}$, respectively. Note that such poles occur for all $J\in\{1,\dots,N\}$, and not just for those such that $s_{IJ}\neq 0$, as we saw for the perturbative poles. For each $\mu=1,2,3$, the integers $a_I(\alpha_\mu)$ and $l_J(\alpha_\mu)$ are computed from some $N$-coloured partition $(\ptt^{(1)},\dots,\ptt^{(N)})$ of total weight $K_\mu$. We follow the notation set out in Appendix \ref{app: 6d conventions}. Namely, $\alpha_\mu$ is a box in the Young tabelaux $Y(\ptt^{(I)})$. $a_I(\alpha_\mu)$ denotes the arm length of $\alpha_\mu$ in $Y(\nu^{(J)})$, while $l_J(\alpha_\mu)$ is its leg length relative to the Young tableaux $Y(\nu^{(J)})$. The only information we will need here is that $|a_I(\alpha_\mu)|,|l_J(\alpha_\mu)| \le K_\mu$.\\

Next, we state the contour prescription for the integral over the $\lambda_I$, as set out in \cite{Kim:2013nva} (see also \cite{Kim:2016usy}). We integrate each $\lambda_I$ over a contour $\Gamma_I(\beta,\epsilon_\alpha,m;L)$ which wraps the cylinder $\lambda_I\sim \lambda_I+2\pi$ precisely once in the right-handed sense. The integrand is meromorphic, and thus the integral is invariant under any continuous deformation of the contour that does not pass over any poles. Thus, we need only specify where the contours sit relative to the various poles. Following \cite{Kim:2013nva}, our strategy is as follows. We first go to a particular point $(\beta,\epsilon_\alpha,m) = (\beta^*,\epsilon_\alpha^*,m^*)$ in parameter space where the positions of the poles simplifies, and we are able to choose the contours $\Gamma_I(\beta^*,\epsilon_\alpha^*,m^*;L)$ quite simply. Then, to arrive at the general contour $\Gamma_I(\beta,\epsilon_\alpha,m;L)$, we continuously vary the parameters as $(\beta^*,\epsilon_\alpha^*,m^*)\to (\beta,\epsilon_\alpha,m)$ while suitably deforming the contour such that no poles pass through it as we do so.

We choose our reference point $(\beta^*,\epsilon_\alpha^*,m^*)$ such that the $\epsilon^*_\alpha$ and $(m^*+\frac{\beta^*}{2L})$ are pure imaginary, and $\text{Re}(\beta^*)>0$. This ensures that all poles lie at points with $\lambda_I-\lambda_J \in \mathbb{R}$. Then, we integrate $\lambda_I$ along the circle defined by\footnote{Note that \cite{Kim:2013nva} dictates that the contour shift should be coordinated with the sign of $\text{Re}(s_I\hat{\beta})$. However, our condition that we should have $\text{Re}(\beta^*)>0$ ensures that this sign matches the sign of $\text{Re}(s_I)$, and so the two shifts are equivalent.}
\begin{align}
  \text{Im}(\lambda_I) = - \zeta s_I
  \label{eq: shifted contours}
\end{align}
for some $\zeta>0$. Such contours avoid all poles (\ref{eq: pert poles}) from the perturbative piece of the integrand. In contrast, whenever we have some $I,J$ with $s_{IJ}=0$, there are still some poles in the non-perturbative piece that are not lifted by the shift by $\zeta$, and thus lie directly on the contours (\ref{eq: shifted contours}). The final contours $\Gamma_I(\beta^*, \epsilon^*_\alpha,m^*;L)$ are defined by an infinitesimal deformation away from (\ref{eq: shifted contours}), whereby the contour is deformed to go either over or under these remaining poles. Precisely whether we go over or under is determined by the values of $a_I(\alpha_\mu)$ and $l_J(\alpha_\mu)$, but we do not need the details of this prescription for our purposes here. Finally, note that the value of the integral is independent of the value of $\zeta>0$, precisely because no poles cross the contours as we vary $\zeta$.

The contours $\Gamma_I(\beta,\epsilon_\alpha,m;L)$ for generic parameters satisfying (\ref{eq: eps constraint}) are then found by continuously varying the parameters to these values, while suitably deforming the initial contours to avoid any poles crossing them. Note, while we chose the contours $\Gamma_I(\beta^*, \epsilon^*_\alpha,m^*;L)$ independently of $L$, the presence of $L$ in the position of the perturbative poles (\ref{eq: pert poles}) means that the $\Gamma_I(\beta,\epsilon_\alpha,m;L)$ will generically depend on $L$.\\

The key observation then is that the finiteness of $(\beta,\epsilon_\alpha,m)$, together with the upper bound on $|a_I(\alpha_\mu)|$ and $|l_J(\alpha_\mu)|$ and the role of $L$ in the position of the perturbative poles (\ref{eq: pert poles}), ensures that the absolute change in the imaginary part of each pole as we flow from $(\beta^*,\epsilon^*_\alpha,m^*)$ to $(\beta,\epsilon_\alpha,m)$ is bounded from above, and further that this bound can be made independent of $L$. It follows, then, that we can choose $\zeta$ sufficiently large (and indeed independently from $L$) such that the deformed contours $\Gamma_{I}(\beta,\epsilon_\alpha,m;L)$ can be chosen to satisfy
\begin{align}
  \text{Im} (\lambda_{I'}) \,\,\left\{\begin{aligned}
  	\,\,&< -\left(\tfrac{1}{3}\text{Re}(\epsilon_+)+1\right) s_{I'}<0\quad \text{if } s_{I'}>0		\\
  	\,\,&> -\left(\tfrac{1}{3}\text{Re}(\epsilon_+)+1\right) s_{I'}>0\quad \text{if } s_{I'}<0	
  \end{aligned}\right. \qquad \text{at all points along } \Gamma_{I'}(\beta,\epsilon_\alpha,m;L)
  \label{eq: contour bound}
\end{align}
where the indices $I',J',\dots$ run over all $I,J,\dots = 1,\dots, N$ such that $s_{I'}\neq 0 $. We can further bound from above the arc length of each $\Gamma_{I}(\beta,\epsilon_\alpha,m;L)$, with a bound that is independent of $L$.

Next, for each such $I'$, let us replace the integration variable $\lambda_{I'}$ with $\rho_{I'}$, defined by
\begin{align}
  \lambda_{I'} = -i s_{I'} \left(\frac{1}{3}\epsilon_++1+\rho_{I'} \right)
\end{align}
Then, the property (\ref{eq: contour bound}) of $\Gamma_{I'}(\beta,\epsilon_\alpha,m;L)$ is rephrased as the statement that at all points along $\Gamma_{I'}(\beta,\epsilon_\alpha,m;L)$, we have $\text{Re}(\rho_{I'})>0$. Then let us study the behaviour of the exponential part of the integrand (\ref{eq: ILKK def}) along the contours $\Gamma_{I}(\beta,\epsilon_\alpha,m;L)$. We have
\begin{align}
  e^{\frac{1}{2}\beta \sum_I s_I^2 }\exp\left(-L\sum_I \left(-\frac{1}{3}\epsilon_+ s_I^2 + i s_I \lambda_I\right)\right)	 &=e^{\frac{1}{2}\beta \sum_{I'} s_{I'}^2 }\exp\left(-L\sum_{I'} \left(-\frac{1}{3}\epsilon_+ s_{I'}^2 + i s_{I'} \lambda_{I'}\right)\right)		\nn\\
  &= \exp \left(-L \sum_{I'} \rho_{I'} s_{I'}^2\right)\exp\left(\left(\frac{1}{2}\beta - L\right)\sum_{I'} s_{I'}^2\right)
\end{align}
and hence, for all points in the integration domain, we have
\begin{align}
  \left|e^{\frac{1}{2}\beta \sum_I s_I^2 }\exp\left(-L\sum_I \left(-\frac{1}{3}\epsilon_+ s_I^2 + i s_I \lambda_I\right)\right)\right| \le \left|\exp\left(\left(\frac{1}{2}\beta - L\right)\sum_I s_I^2\right)\right|
\end{align}
Then, the right-hand-side vanishes exponentially equickly as $L\to\infty$ for all $s\neq (0,\dots,0)$. Finally, the $L$-independent upper bound on the arc length of each $\Gamma_{I}(\beta,\epsilon_\alpha,m;L)$ ensures an $L$-independent upper bound on the volume of the integration domain. Thus, we have
\begin{align}
  \lim_{L\to\infty} \hI^{(s)}_{L;K_1,K_2,K_3}(\beta,\epsilon_\alpha,m) = 0
\end{align}
for all $s\neq (0,\dots,0)$ as required.


\subsection{Collapse of the zero flux term}\label{subsec: zero flux}

We are left to determine the limit of the zero flux term,
\begin{align}
  \mathcal{K}(\beta,\epsilon_\alpha,m):&=\lim_{L\to\infty} \hI^{(s=(0,\dots,0))}_L\!\left(\hat{\beta}=\tfrac{2}{3}\epsilon_+ + \tfrac{\beta}{L},\,\hat{v}_\alpha=\epsilon_\alpha - \tfrac{2}{3}\epsilon_+,\,\hat{m}=m\right)	
\end{align}
This is straightforward; the fact that $\lim_{\beta\to + \infty} \Z_\text{inst}(\beta,\epsilon_1,\epsilon_2,m,\lambda_I) = 1$ together with the conditions (\ref{eq: eps constraint}) ensures that the first two factors of $\Z_\text{inst}$ appearing in $\Z_\text{non-pert}^{(s=(0,\dots,0))}$ go to unity in the limit. Using also (\ref{eq: perturbative zero flux}), we find
\begin{align}
  \mathcal{K}(\beta,\epsilon_\alpha,m)& = \int d\mu_{\text{Haar}}[\lambda]\, \Z_{\text{inst}}(\beta,\epsilon_1, \epsilon_2,m, \lambda_I)
  \label{eq: K result}
\end{align}
matching precisely the form of $\I(\beta,\epsilon_\alpha,m)$ in (\ref{eq: LC index QM result}). Let us finally derive the contours $\Gamma_I(\beta,\epsilon_\alpha,m)$ over which the integrals in (\ref{eq: K result}) are taken. The algorithm of \cite{Kim:2013nva,Kim:2016usy} tells us to first choose $\epsilon_1,\epsilon_2$ pure imaginary, which places all poles at points with $\lambda_I - \lambda_J \in \mathbb{R}$. The contours chosen then lie simply along the real line from $0$ to $2\pi$, with small deformations either upwards or downwards in order to avoid these poles. The prescription by which to do this states that we should simply leave the contours on the real line, and shift $\epsilon_\alpha \to \epsilon_\alpha + \eta$ for $\eta$ a small, positive real number. This has the effect of moving some poles upwards into the upper-half-plane, and some downwards into the lower-half-plane. Then, since we are only interested in values of $\mathcal{K}(\beta,\epsilon_\alpha,m)$ with $\epsilon_\alpha$ satisfying (\ref{eq: eps constraint}), we learn that for all such $\epsilon_\alpha$, the contours $\Gamma_I(\beta,\epsilon_\alpha,m)$ can be chosen simply to lie along the real line from $0$ to $2\pi$, matching precisely our prescription for $\I(\beta,\epsilon_\alpha,m)$. Hence, we find
\begin{align}
  \lim_{L\to\infty} \hI^{(s=(0,\dots,0))}_L\!\left(\hat{\beta}=\tfrac{2}{3}\epsilon_+ + \tfrac{\beta}{L},\,\hat{v}_\alpha=\epsilon_\alpha - \tfrac{2}{3}\epsilon_+,\,\hat{m}=m\right)	= \I(\beta,\epsilon_\alpha,m)
\end{align}
as required.


\bibliographystyle{JHEP}
\bibliography{paper}

\end{document}